\documentclass[twocolumn]{aastex631}
\usepackage{amsmath}
\usepackage{subfigure}
\usepackage{xcolor}
\usepackage{float,color}
\usepackage{gensymb}
\usepackage{placeins}
\usepackage{appendix}
\usepackage{cancel}
\usepackage{comment}

\usepackage{mdframed}
\usepackage{booktabs}

\def\ninja{\texttt{NINJA} }
\def\astrid{\texttt{Astrid} }
\def\bluetides{\texttt{Bluetides} }
\def\illustris{\texttt{Illustris} }
\def\tng{\texttt{TNG} }
\def\tngh{\texttt{TNG-100} }
\def\tngt{\texttt{TNG-300} }
\def\mbi{\texttt{MB-I} }
\def\mbii{\texttt{MB-II} }
\def\massiveblack{\texttt{MassiveBlack} }
\def\Lbox{L_{\textrm{box}}}
\def\Npart{N_{\textrm{part}}}
\def\Mbh{M_{\textrm{BH}}}
\def\Lbol{L_{\textrm{bol}}}
\def\Lsun{L_{\odot}}

\def\mdm{m_{\textrm{dm}}}
\def\mgas{m_{\textrm{gas}}}
\def\mstar{m_{\textrm{star}}}
\def\msun{M_{\odot}}

\renewcommand{\software}[1]{{\small #1}}
\def\mpgadget{\software{MP-GADGET}\,\,}
\def\pgadget{\software{PGADGET-3}\,\,}
\def\treepm{\software{TreePM}\,\,}
\def\rockstar{\software{ROCKSTAR}\,\,}

\begin{document}

\title{Forecasting properties of detectable massive binary black hole mergers in the era of space based gravitational-wave detectors}

\author[0000-0002-0870-2993]{Sourabh Magare}
\affiliation{Inter-University Centre for Astronomy and Astrophysics, Post Bag 4, Ganeshkhind, Pune - 411007, India}

\author[0009-0004-5627-7196]{Abhinav Roy}
\affiliation{School of Physical Sciences, National Institute of Science Education and Research, Jatni 752050, Odisha, India}
\affiliation{Homi Bhabha National Institute, Training School Complex, Anushaktinagar, Mumbai 400094, Maharashtra, India}

\author[0000-0001-5318-1253]{Shasvath J. Kapadia}
\affiliation{Inter-University Centre for Astronomy and Astrophysics, Post Bag 4, Ganeshkhind, Pune - 411007, India}

\author[0000-0003-3081-0189]{Nishikanta Khandai}
\affiliation{School of Physical Sciences, National Institute of Science Education and Research, Jatni 752050, Odisha, India}
\affiliation{Homi Bhabha National Institute, Training School Complex, Anushaktinagar, Mumbai 400094, Maharashtra, India}

\author[0000-0002-9062-1921]{R. Srianand}
\affiliation{Inter-University Centre for Astronomy and Astrophysics, Post Bag 4, Ganeshkhind, Pune - 411007, India}

\begin{abstract}
Gravitational waves (GWs) from massive black hole (MBH) mergers will provide a novel way to probe the high-redshift universe and are key to understanding galactic dynamics and evolution. In this work, we analyze MBH mergers, their GW signals and detectability, as well as their population properties, using the cosmological hydrodynamical simulation - \ninja Simulation Suite. 
We discuss the effect of resolution and finite volume on the black hole mass function (BHMF), which in turn 
limits the mergers associated with low mass black holes, $\Mbh \lesssim 10^{6.5} \msun$. We find the upper limit on the total mass of the MBH binaries detectable by LISA to be $\sim 10^{8.4}\msun$. We also find that adding time delays pertaining to dissipative processes like dynamical friction and stellar hardening during the final stages of the inspiral for which the simulation lacks sufficient resolution to model, considerably shifts the peak of redshift distribution of detectable binaries from $z\sim0.5$ to $z\sim0.1$. Time delays reduce the number of detectable GW events but on the other hand their signal-to-noise is increased. From the observational point of view, we find a strong correlation between the SFR and $L_{\rm bol}$ at high redshifts for the detectable LISA binaries. This may prove to be a future application in the coincident observation of MBH binaries by GW and electromagnetic observations. 

\end{abstract}

\section{Introduction}

Gravitational waves (GWs) emitted by the merger of stellar-mass binary black holes (BBHs) are being routinely detected by the LIGO-Virgo-Kagra (LVK) network of GW detectors \citep{2015CQGra..32g4001L, 2015CQGra..32b4001A, 2021PTEP.2021eA101A}, currently in its fourth observing run (O4). So far, over $200$ BBHs have been observed \citep{2025arXiv250904348T}, shedding light on stellar-mass BH populations \citep{2025arXiv250818083T}, providing unique tests of General Relativity (GR) in the strong-field regime, as well as stringent constraints on alternative theories of gravity \citep{2021arXiv211206861T}.

However, the LVK detector network has a limited bandwidth, spanning $\mathcal{O}(10-1000 {\rm Hz})$, with a ``seismic wall'' that prohibits frequencies less than $10$ Hz from being observed. This, in turn, limits the mass-range of detectable compact binary coalescences (CBCs) to  $\sim 1 -100 M_{\odot}$ \citep{2019Natur.568..469M}. Since to the first approximation,
the frequency of GWs is inversely proportional to the total mass of the compact objects, mergers of more massive black holes emit GWs in the milli-hertz
range \citep{2019ApJ...883L..27M,2021hgwa.book.....B}. The space-based Laser Interferometer Space Antenna (LISA), expected to be operational in the near future, will
have a bandwidth $100 \mu$Hz to $100$mHz and thus is expected to detect the merger of massive black hole
binaries (MBHBs) in the mass range of $\sim$ $10^{4} - 10^{8} \msun$ \citep[e.g][]{2017arXiv170200786A, 2021hgwa.book.....B}.
These black holes (BHs) reside at the centers of most galaxies and thus play an important role in the evolution of their hosts \citep{1995ARA&A..33..581K,1998AJ....115.2285M,2013ARA&A..51..511K,2015ApJ...813...82R,2019ApJ...887..245S}. 

Further, the coincident detection of MBHBs in GWs and by EM telescopes has the potential to provide unprecedented understanding of the
evolution of massive black holes (MBHs) \citep{2010ApJ...709..774K,2019Galax...7...63G}. These complementary information will allow us to discover the seeds of
MBHs \citep{2007MNRAS.377.1711S} and possibly obtain information about their environment at high redshifts \citep{2019MNRAS.486.4044B}.
Since, GWs can directly measure the luminosity distance of the source, and from the EM sources we can get the redshift information,
their combination can be used as a probe of expansion rate of the universe at high redshifts \citep{1986Natur.323..310S,2005ApJ...629...15H}.
Coincident observations of MBHBs in EM and GW can also be used to understand the properties of gas in the accretion disk \citep{2019MNRAS.486.2754D},
to provide a tighter constraint on the $M-\sigma$ relation, to better understand the initial seeding mechanism 
\citep{2009MNRAS.400.1911V},
and to understand dynamical evolution of MBHs and MBHBs in general \citep{2015MNRAS.452.2337K}. 

According to the $\Lambda\textrm CDM$ model, dark matter haloes and the galaxies they host form through hierarchical merging. As a result, MBHs form at the centers of galaxies, and the most natural way to bring pairs of MBHs into close proximity is via galaxy mergers. 
MBHs are also believed to accrete gas from their surroundings and power
Active Galactic Nuclei (AGN) and quasars \citep{1982MNRAS.200..115S,2002MNRAS.335..965Y}. These in turn exert feedback on the
galaxies and can quench its star formation \citep{2005Natur.433..604D}. Although MBHs have an important role in galaxy evolution, the mechanisms that regulate the growth of MBH and the physics surrounding the AGN feedback still have
lot of gray areas. Scaling relations between the MBH and the host galaxies -- e.g. mass of the MBH and dispersion velocity of
galaxy \citep{2000ApJ...539L...9F,2009ApJ...698..198G} --  are proof of the synergic evolution of the MBH and it's host galaxy. However, these relations are found to be more complicated than previously thought. Moreover, MBHs can also be primordial, produced during the very early universe. LISA can provide us with accurate estimates of masses, spins and
orbital parameters of MBHBs, which could enable unique constraints on the formation and growth of MBHs.

Studies to estimate the merger rates and properties of the MBHBs population have been going on for many years and can be broadly
classified into types- $(1)$ using analytical models, $(2)$ using semi-analytical models and $(3)$ by cosmological hydrodynamical
simulations which self-consistently evolve the properties of MBHs and their host galaxies.
In order to obtain MBH merger rates, 
analytical and semi-analytical approaches rely on halo merger-trees, based on the extended Press-Schecter theory 
\cite{1974ApJ...187..425P,2006MNRAS.371.1992E,2008MNRAS.383..557P} or from N-body simulations \citep{2007MNRAS.380.1533M}.
These approaches can be highly flexible and use increasingly complicated models calibrated 
on cosmological hydrodynamical simulations and observations, and require a relatively low computational cost \citep{2020ApJ...904...16B}. 

On the other hand, cosmological hydrodynamical simulations \citep{2012ApJ...745L..29D, 2014Natur.509..177V, 2014MNRAS.442.2304H, 
2015MNRAS.446..521S, 2015MNRAS.450.1349K, 2016MNRAS.455.2778F, 2016MNRAS.460.2979V, 2018MNRAS.473.4077P, 2019MNRAS.486.2827D, 2025arXiv251013976Z} have spatial information of the galaxies, can reach
high complexities but are computationally expensive and limited in dynamic range, i.e., scales where the GW signal from MBHB mergers becomes important. In recent years, large-scale cosmological simulations, which dynamically incorporate galaxy and MBH evolution and feedback processes, are emerging as an ever more essential resource for probing and refining the physics of galaxy formation because of swift advancements in computing capabilities. The cosmological hydrodynamical simulations offer an ideal setting for studying the evolution and mergers of MBHs. Additionally, cosmological hydrodynamical simulations possess the statistical strength to forecast merger rates ahead of forthcoming observations. These simulations have been used 
to predict the GW signal from MBH mergers \citep{2016MNRAS.463..870S, 2017MNRAS.464.3131K, 2020MNRAS.491.2301K, 2022MNRAS.511.5241S, 2022ApJ...933..104L, 2024MNRAS.531.1931K, 2024MNRAS.52711766D, 2025ApJ...988L..74ZZ,2025ApJ...991L..19C, 2025ApJ...993..199W}. 

In this work, we make predictions about the detectability of GWs from MBHBs in LISA using the cosmological hydrodynamical simulation suite \ninja. Recently, the GW signal was estimated \citep{2025ApJ...993..199W} for LISA 
from the \astrid simulation \citep{2025arXiv251013976Z}, but time delays in the MBHB merger events, 
which we consider in this work, have been ignored. Our results are consistent, but complement the results of 
\cite{2025ApJ...993..199W}. We discuss the role of time delays in reducing the number of GW events
but at the same time enhancing the GW signal. We also discuss the importance of convergence 
in the MBH properties, which one has to consider to reliably estimate the GW signal.

\section{Methodology}
In this section, we present the cosmological hydrodynamical simulations, which are post-processed to construct MBH merger histories on which the GW signal is modeled.

\subsection{The \ninja Simulation Suite}
We describe the low-redshift runs of the
\ninja (\textbf{\texttt{N}}ISER-\textbf{\texttt{I}}UCAA \textbf{\texttt{N}}ew Simulations of \textbf{\texttt{J}}WST G\textbf{\texttt{A}}laxies) \citep{Behera2025}
simulation suite used for this work. The \ninja simulations is a suite of cosmological hydrodynamical simulations
currently being run at NISER\footnote{National Institute of Science Education and Research (\href{www.niser.ac.in}{www.niser.ac.in})}
and IUCAA\footnote{Inter-University Centre for Astronomy and Astrophysics(\href{www.iucaa.in}{www.iucaa.in})}
to better understand galaxy formation at high redshift as probed by
the James Webb Space Telescope (JWST)\footnote{\href{www.jwst.nasa.gov}{www.jwst.nasa.gov}}. 
The high-redshift \ninja runs with $\Npart = 2\times 2040^3$ particles in comoving boxes of side
$\Lbox = 50 - 250 \,\,\textrm{Mpc/h}$ vary the subgrid models of star formation, black hole growth, and associated feedback processes
to better understand these recent observations, which are currently restricted to $z \gtrsim 5$.
Here we present the low-redshift \ninja runs  with $\Npart = 2\times 1008^3 - 2\times 640^3$ particles in cosmological volumes
similar to the high-redshift runs, but the galaxy formation parameters are similar to the \astrid \citep{2022MNRAS.512.3703B,2022MNRAS.513..670N,2023MNRAS.522.1895C, 2025arXiv251013976Z}
and the fiducial \illustris-\tng \citep{2017MNRAS.465.3291W,2018MNRAS.473.4077P}
simulations.
The cosmological parameters for the low-redshift runs are $(\Omega_m,\Omega_\Lambda,\Omega_b,h,\sigma_8,n_s) = (0.2814, 0.7186, 0.0464, 0.697, 0.81, 0.971)$.
These are the same as the ones used in \bluetides \citep{2016MNRAS.455.2778F} and
consistent with the recent results from the Dark Energy Spectroscopic Instrument \citep[DESI,][]{2025JCAP...02..021A}\footnote{https://www.desi.lbl.gov/} 
in conjunction with Big Bang Nucleosynthesis \citep[BBN,][]{2024JCAP...06..006S} priors, assuming a flat cosmology.
Here $\Omega_m, \Omega_\Lambda = 1-\Omega_m,\Omega_b$ are the cosmological density parameters associated with non-relativistic matter, the cosmological constant
and baryons. $h$ is the dimensionless Hubble parameter, $\sigma_8$ is the mass variance at smoothed on 8 megaparsec scales $z=0$
and is used to normalize the matter power spectrum, and $n_s$ is the primordial spectral index of matter fluctuations.

\begin{table*}[t]
\centering
\begin{tabular*}{\textwidth}{@{\extracolsep{\fill}}cccccccc}
\toprule
\ninja & $L_{\mathrm{box}}$ & $N_{\mathrm{part}}$ & $\mdm$ & $\mgas$ & $\mstar$ & $\epsilon$ & $z_{\mathrm{final}}$ \\
Run & $\left(h^{-1}\mathrm{Mpc}\right)$ &  & $\left(h^{-1}\msun\right)$ & $\left(h^{-1}\msun\right)$
& $\left(h^{-1}\msun\right)$ & $\left(h^{-1}\mathrm{kpc}\right)$ & \\
\midrule

L140\_N1008 & 140 & $2\times 1008^3$ & $1.8\times 10^8$ & $3.5\times 10^7$ & $8.6\times 10^6$ & 4.6 & 0 \\
(fiducial) \\
\midrule

L140\_N896 & 140 & $2\times 896^3$ & $2.5\times 10^8$ & $4.9\times 10^7$ & $1.2\times 10^7$ & 5.2 & 0 \\
(intermediate resolution) \\
\midrule

L140\_N700 & 140 & $2\times 700^3$ & $5.2\times 10^8$ & $1.0\times 10^8$ & $2.6\times 10^7$ & 6.7 & 0 \\
(low resolution) \\
\midrule

L50\_N640 & 50 & $2\times 640^3$ & $3.1\times 10^7$ & $6.1\times 10^6$ & $1.5\times 10^6$ & 2.6 & 0 \\
(high resolution) \\
\midrule

L50\_N1008 & 50 & $2\times 1008^3$ & $7.9\times 10^6$ & $1.5\times 10^6$ & $3.9\times 10^5$ & 1.6 & 0.6 \\
(highest resolution) \\
\bottomrule
\end{tabular*}

\caption{Basic simulation parameters. The columns list the run name, simulation box size
$L_{\mathrm{box}}$, the initial number of particles ($N_{\mathrm{part}}$), the dark matter particle mass $\mdm$,
the gas particle mass $\mgas$, the star particle mass $\mstar \approx \mgas/4$, the gravitational softening length $\epsilon$,
and the final redshift reached by the simulation $z_{\mathrm{final}}$.
All length scales are in comoving units. The initial conditions for the L140 series
are generated with the same seed.}
\label{table_simparam}
\end{table*}

\begin{figure*}[t]
\centering
\includegraphics[width=1\textwidth]{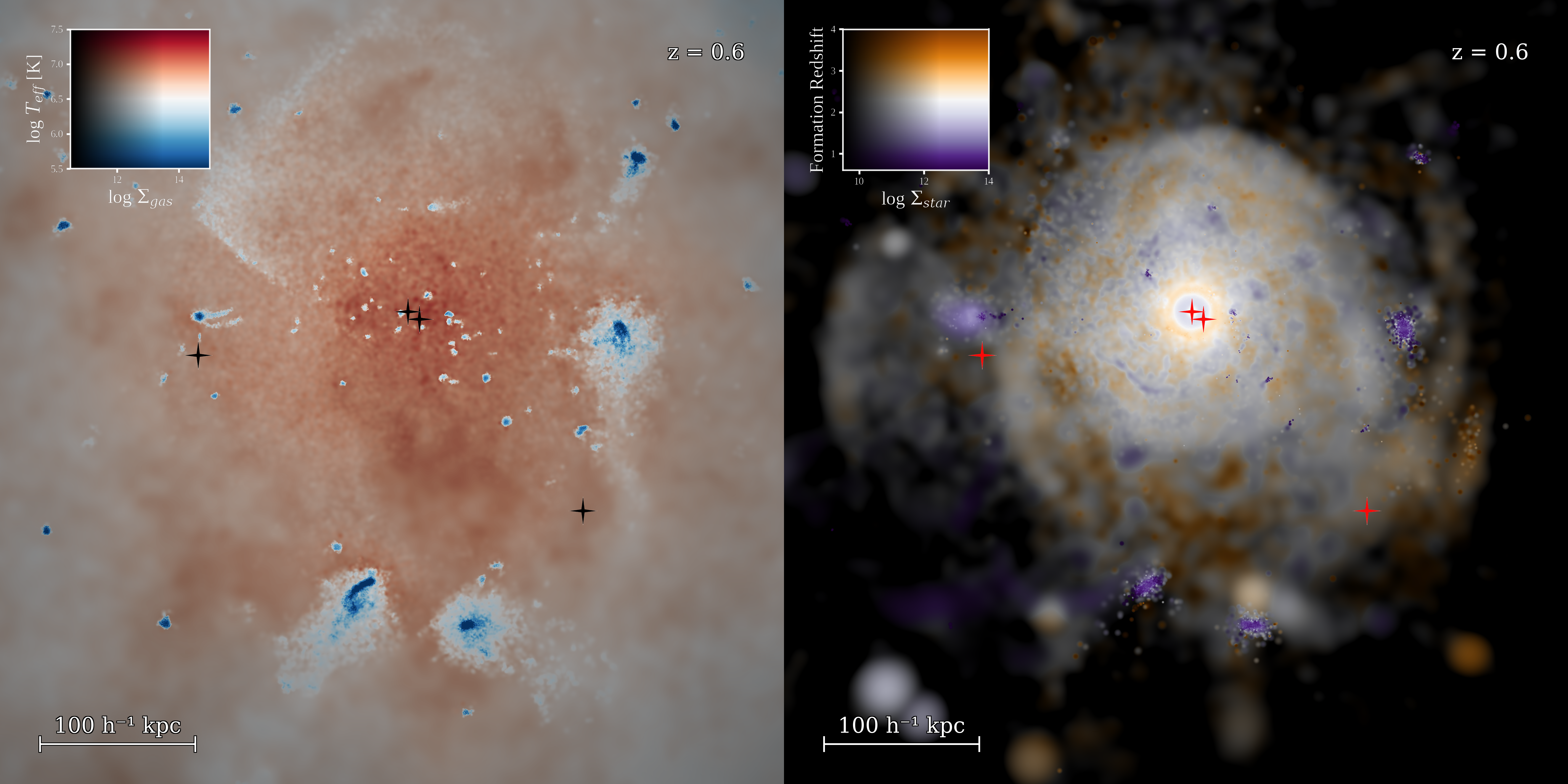}
\caption{Projected gas (left) and stellar (right) properties from the NINJA L50\_N1008 simulation at redshift $z = 0.6$, showing a $500~h^{-1}kpc$ deep projection focused on a halo. 
\textit{Left panel:} Gas surface density ($\log \Sigma_{\rm gas}$ [$M_\odot~hMpc^{-2}$]) color-coded by temperature ($\log T_{\rm eff}$ [K]), revealing the multi-phase structure of the interstellar and circumgalactic medium. 
\textit{Right panel:} Stellar surface density ($\log \Sigma_{\rm star}$ [$M_\odot~hMpc^{-2}$]) color-coded by formation redshift, illustrating the stellar mass distribution and age structure. 
The scale bar indicates $100~h^{-1}kpc$.
Black holes with $M_\bullet>10^{6.5}h^{-1}M_\odot$ in the region are marked with stars (black crosses in the left panel, red crosses in the right panel).
Note the correspondence between cold, clumpy gas patches (blue regions in the left panel) and recent star formation (purple regions in the right panel), particularly along the spiral arms, while multiple black holes near the halo center are in the process of inspiraling and should eventually merge.
}
\label{fig:halo_vis}
\end{figure*}

\ninja has been run with the \mpgadget \citep{2018zndo...1451799F}
code, a massively parallel cosmological
hydrodynamical simulation code, used to study the formation and evolution of halos,
galaxies and MBHs and their interplay with
the matter around and in between them in
cosmological volumes.
\mpgadget is based on \pgadget \citep{2016MNRAS.455.2778F}.
\pgadget was used to run previous large simulation campaigns like the \massiveblack \texttt{I} and \texttt{II} simulations (\mbi, \mbii)
\citep{2012ApJ...745L..29D,2015MNRAS.450.1349K}.  
Currently the largest cosmological hydrodynamical simulations like \bluetides \citep{2016MNRAS.455.2778F}
and \astrid \citep{2022MNRAS.512.3703B,2022MNRAS.513..670N} have been run with \mpgadget.

The gravity solver in \mpgadget uses the \treepm \citep{2002JApA...23..185B} algorithm
and the smoothed particle hydrodynamics (SPH) solver adopts the pressure entropy formulation of \cite{2013MNRAS.428.2840H}.
Most of the applied subgrid model parameters are the same as  
\astrid \citep{2022MNRAS.512.3703B,2022MNRAS.513..670N,2023MNRAS.522.1895C} with some differences in the black hole model,
which are highlighted. 
A multiphase star formation model is implemented as \citep{2003MNRAS.339..289S}.
Gas cools radiatively  \citep{1996ApJS..105...19K} and by metal-line cooling \citep{2014MNRAS.444.1518V}. 
Self-shielding correction in dense regions is done for neutral hydrogen based on the prescription of 
\cite{2013MNRAS.430.2427R}. Star formation is also modified
to account for the effect of molecular hydrogen 
which is estimated based on the local column density and metallicity \citep{2011ApJ...729...36K}.
Massive stars and supernovae enrich the surrounding medium by
returning mass and metals to neighboring gas particles \citep{2013MNRAS.436.3031V,2018MNRAS.473.4077P}.
A variable kinetic wind feedback due to supernovae which depends on the local dark matter velocity dispersion is implemented
as in   \cite{2010MNRAS.406..208O,2013MNRAS.436.3031V}.

Black hole growth is modeled in the
same way as in the \mbi and \mbii simulations  based on the black hole sub-grid model developed in \cite{2005Natur.433..604D,2005MNRAS.361..776S} and
first implemented in cosmological simulations in \cite{2008ApJ...676...33D}.
A MBH with seed mass $\Mbh^{\textrm{seed}} = 5 \times 10^5 \msun/\textrm{h}$ is placed in a newly formed on-the-fly friends of friends (FOF) \citep{1985ApJ...292..371D} halo
of mass $M_{\textrm{halo}}^{\textrm{seed}} = 10^{10}\msun/\textrm{h}$ halo if it does not already contain a MBH. 
We also add an additional minimum stellar mass criterion to the halo for seeding black holes, $M_{\textrm{star}}^{\textrm{seed}} = 5\times10^{6} \msun/\textrm{h}$.
For the L140 series this additional condition translates to seeding halos with a minumum of 1 star particle and 
for the L50 box it is 4 star particles. As discussed later, at low redshifts and for black holes with masses $\log_{10}(\Mbh/\msun) \gtrsim  6.5$
this additional stellar mass threshold for seeding black holes does not play an important role.

A constant seed MBHs mass has been employed
in \mbi, \mbii, \bluetides and the \tng simulations with similar seed MBH and halo masses.
\astrid on the other hand, has implemented a variable MBH seed mass in the range
$\Mbh^{\textrm{seed}} = 3\times10^4 - 3\times10^5 \msun/\textrm{h}$ drawn randomly from a powerlaw mass distribution with index $n=-1$ in FOF halos with $M_{\textrm{halo}}^{\textrm{seed}} = 10^{10}\msun/\textrm{h}$
due to their better mass resolution.
MBHs accrete gas at a rate given by the Bondi-Hoyle-Lyttleton accretion rate \citep{1939PCPS...35..405H,1944MNRAS.104..273B,1939PCPS...35..405H}
\begin{equation}
\dot{M}_{\textrm{BH}} = \alpha\frac{4\pi G^2\Mbh^2 \rho}{\left(c_{\textrm{s}}^2 + v_{\textrm{rel}}^2\right)^{3/2}}
\label{eq_bondi_accretion}
\end{equation}
The parameter $\alpha = 100$ is used to boost the accretion rate, which is underestimated due to
the unresolved interstellar medium. $\Mbh$ is the black hole mass, $\rho, c_{\textrm{s}}, v_{\textrm{rel}}$ are
the local gas density, sound speed, and the relative velocity of the MBH  with respect to the surrounding gas, and $G$ is the gravitational constant.
We allow for super-Eddington accretion but cap the accretion at twice the Eddington rate
\begin{equation}
\dot{M}_{\textrm{Edd}} = \frac{4\pi G \Mbh m_{\textrm{p}}}{\eta \sigma_{\textrm{T}}c}
\label{eq_eddington_rate} .
\end{equation}
Here, $c$ is the speed of light, $\sigma_{\textrm{T}}$ is the Thomson cross section, and $m_{\textrm{p}}$ is the mass of the proton.
$\eta = 0.1$ is the radiative efficiency, i.e. a mass-to-light conversion efficiency in an accretion disk  according to \cite{1973A&A....24..337S}.
The MBH radiates with a bolometric luminosity: $\Lbol = \eta \dot{M}_{\textrm{bh}} c^2$.

\begin{figure}[h]
\centering
\includegraphics[width=1\columnwidth]{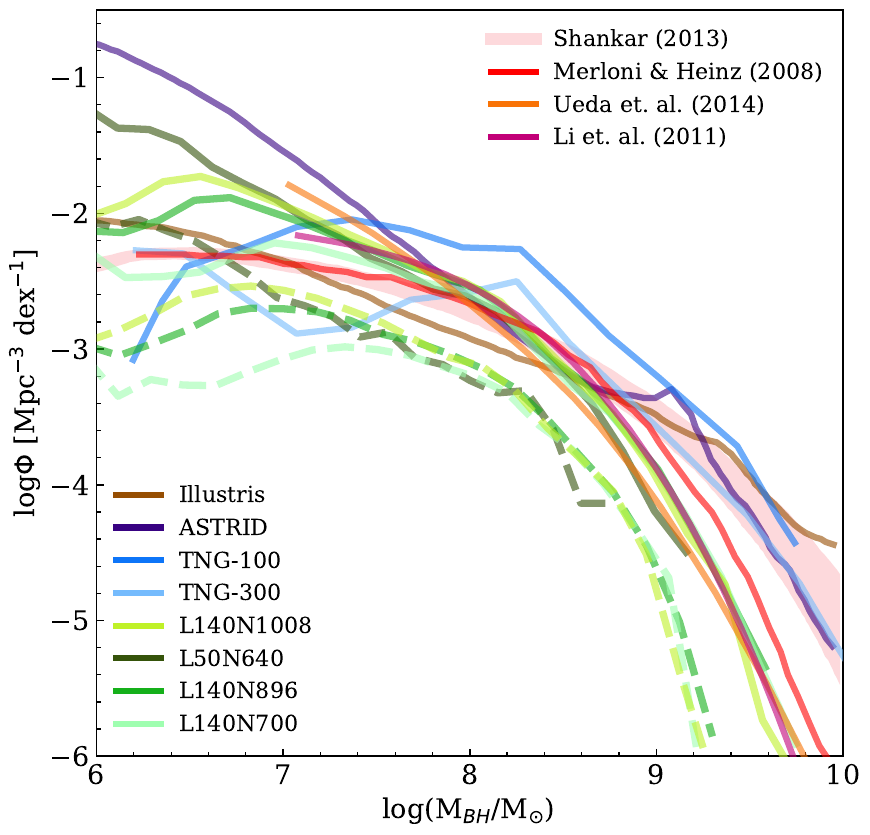}
\caption{Comparison of the black hole mass function for different simulation volumes of \ninja with other simulations and observations at $z=0$.
The green lines represent the results of the \ninja runs (solid for centrals and dashed for satellites), the solid blue, brown and purple lines represent the results from \tng \citep{2021MNRAS.503.1940H},  \illustris \citep{2015MNRAS.452..575S}, and the \astrid \citep{2025ApJ...993..199W} simulations.
The pink shaded region and the red, orange and magenta lines are observational constraints from \cite{2013CQGra..30x4001S}, \cite{10.1111/j.1365-2966.2008.13472.x}, \cite{2014ApJ...786..104U}, and \cite{Li_2011} respectively.
The error estimates from  \cite{10.1111/j.1365-2966.2008.13472.x} are extremely small and have not been included. The pink shaded region of \cite{2013CQGra..30x4001S} represents the $1\sigma$ uncertainty of the mass function assuming the $\Mbh-\sigma$ relation from \citet{2013ApJ...764..184M}.}
\label{fig_bhmf}
\end{figure}

Feedback from MBH switches between thermal (due to high accretion rate) and kinetic mode (low accretion rate) depending on their accretion
rate relative to the Eddington rate.
\mbi and \mbii had only thermal mode feedback, where 5\% of the radiated energy couples thermally to the surrounding gas.
This coupling is the same as in \astrid and we consider the same value in \ninja. Kinetic feedback is included in \astrid for $z<2.4$ \citep{2023MNRAS.522.1895C}
and follows the \tng model \citep{2017MNRAS.465.3291W}. The high accretion mode is defined if $(\dot{M}_{\textrm{BH}}/\dot{M}_{\textrm{Edd}}) \geq \chi$ (low accretion otherwise),
where $\chi$ is a MBH mass-dependent function capped at 0.1
\begin{equation}
\chi = \textrm{min}\left[\chi_0\left(\frac{\Mbh}{10^8\msun}\right)^{\beta},0.1  \right] \quad \chi_0 = 0.002 \,\,, \beta = 0.2
\end{equation}
Kinetic feedback is particularly important at low redshifts, and the above parameters are chosen so that massive galaxies
get quenched rapidly. Indeed \mbii had an excess of massive galaxies at $z=0$ compared to the
estimates of the galaxy stellar mass function due to inadequate quenching.
The particular MBH mass scale above which kinetic feedback becomes important is at
$\log_{10}(\Mbh/\msun) \approx 8.9 $. Note that the local environment, redshift, and accretion history are particularly important
in determining if the MBH has entered a kinetic feedback-dominated regime or not.
We typically find that this sets in at $z < 2$ for MBH  $\log_{10}(\Mbh/\msun) \gtrsim 8 $ \citep{abhinav_msc_thesis}.
Similar mass MBHs at higher redshift and low mass MBHs across all redshifts do not accrete in the low accretion mode.   
Coupling efficiency in low accretion, kinetic mode is a density dependent value capped at 0.2.
The energy injection mechanism is similar to feedback from supernovae and depends on the local
dark matter velocity dispersion, but occurs in bursts whenever a minimum energy injection threshold is reached.
We choose the parameter controlling this threshold to be the same as the fiducial \tng model \citep{2017MNRAS.465.3291W}
which is 4 times larger than the value adopted in \astrid \citep{2023MNRAS.522.1895C}.  

In \pgadget, the most MBHs were pinned, by hand, to the halo center, whenever they got a force update, i.e., at every active timestep.
\mpgadget alleviates this issue of MBHs straying from the halo center, by implementing a subgrid dynamical friction model \citep{1943ApJ....97..255C} 
due to collisionless particles (like stars and dark matter) \citep{2022MNRAS.510..531C}, which affects the dynamics of the MBH in a self-consistent manner. 
MBHs also feel a drag force due to surrounding gas \citep{1999ApJ...513..252O}, but this is
typically smaller than the force due to dynamical friction \citep{2022MNRAS.510..531C}. In \pgadget, a pair of MBHs merge if their separation is within
the spatial resolution of the simulation and their relative velocities are smaller than the local sound speed. In \mpgadget the second criterion is
changed to a binding criterion of the pair using relative positions, velocities, and accelerations \citep{2022MNRAS.513..670N}.

We use \rockstar \citep{2013ApJ...762..109B} on dark matter particles to identify substructures in FOF halos. \rockstar also identifies substructures in substructures
recursively. We restrict our definition of substructures to only the first level of the substructure tree. The most massive substructure is identified as the central subhalo 
and the rest as satellite subhalos. We then spatially associate gas, star, and blackhole particles with a subhalo  (central or satellite) so that they
are bound within the virial radius of the associated subhalo. We loosely refer to these subhalos (central and satellites) as halos (or galaxies when discussing the stellar mass) in
the rest of the paper. The center of the halo is defined as the location of the particle with the least
gravitational potential. We associate a single MBH with a galaxy by looking at the black hole closest to the center of the halo within a tenth
of the virial radius. We refer to these  MBHs as central or satellite black holes if their associated halo is a central or satellite halo.
We have also considered different assignments of black holes to halos. For example, if we consider
the most massive black hole within a tenth or fifth of the virial radius, we find little change in the black hole mass function (BHMF)
of the \ninja runs in figure~\ref{fig_bhmf}, for centrals and satellites, for masses $\log_{10}(\Mbh/\msun) \gtrsim 6.5$.
Our method of assigning a single MBH to a single halo (substructure) ignores many other \emph{wandering} blackholes \citep{2023MNRAS.520.3955W,2023MNRAS.525.1479D}
within the halo but without an associated host. Typically, these wandering black holes, relatively low in mass,  lose their host prior to or during a merger.
This typically happens if the host halo is poorly resolved or has a small concentration parameter and is tidally disrupted prior to or during a merger.

For this work, we present five simulations, which are listed in the table~\ref{table_simparam}.
The fifth and most recent run, L50\_N1008, has recently reached $z=0.6$ and is included only for a visualization example in figure~\ref{fig:halo_vis} but is not used in the subsequent analysis as it has not yet evolved to $z=0$.
All the runs have identical cosmological and galaxy formation parameters but differ only in the number of particles and volume.
The L140 series have the same initial conditions. 

We discuss the effect of resolution and finite volume on the BHMF as shown in the figure~\ref{fig_bhmf}.
The fiducial run used for the analysis is L140\_N1008, which is run with $\Npart = 2\times 1008^3$ particles in a comoving box of $\Lbox = 140 \textrm{Mpc/h}$.
The high resolution run here, L50\_N640, has a mass resolution $\sim 17\times$ better compared to the lowest one, L140\_N700, and $\sim 6\times$ better compared to the fiducial run, L140\_N1008.
In comparison, \astrid has a mass resolution which is $\sim 5\times$ and $\sim 27\times$ better than the highest resolution run and the fiducial runs, respectively.

In figure~\ref{fig_bhmf} we look for convergence of the BHMF in the \ninja runs at $z=0$
and also compare them with the BHMF from the simulations of \astrid \citep{2022MNRAS.512.3703B,2022MNRAS.513..670N,2025ApJ...993..199W}, \illustris \citep{2015MNRAS.452..575S},
\tngh, \tngt \citep{2018MNRAS.480.5113M,2018MNRAS.477.1206N,2018MNRAS.475..624N,2018MNRAS.475..648P,2018MNRAS.475..676S,2021MNRAS.503.1940H}
as well as observational constraints of \cite{10.1111/j.1365-2966.2008.13472.x}, \cite{Li_2011},
\cite{2013CQGra..30x4001S} and \cite{2014ApJ...786..104U}. 

\cite{2013CQGra..30x4001S} and \cite{Li_2011}  assume that every galaxy hosts one black hole 
and convert galaxy abundances into black hole abundances using a scaling relation. 
\cite{2013CQGra..30x4001S} uses $\Mbh-\sigma$  
scaling relation of \cite{2013ApJ...764..184M} for early type galaxies,
to estimate the BHMF, where $\sigma$ is the stellar velocity dispersion. 
\cite{Li_2011} on the other hand use $\Mbh-M_{\textrm{sph}}$ relation for their estimates of the BHMF, where
$M_{\textrm{sph}}$ is the host spheroid mass. 

\cite{10.1111/j.1365-2966.2008.13472.x} and \cite{2014ApJ...786..104U} on the other hand 
use the argument of \cite{1982MNRAS.200..115S}, i.e., the growth of a black hole is mainly from accretion,
coupled with the observed x-ray luminosity (XLF) function in conjunction with the black hole continuity equation
to obtain the BHMF. These calculations involve an assumed radiative efficiency $\epsilon_{\textrm{rad}} \sim 10\% $
and an average Eddington ratio. The results of \cite{10.1111/j.1365-2966.2008.13472.x} 
use the XLF of \cite{2008ApJ...679..118S}, based on a sample of $\sim 700$ AGNs. \cite{2014ApJ...786..104U}
on the other hand use a $\sim 6\times$ larger sample of AGNs to obtain the XLF and the BHMF. Both observations
obtain the XLF out to $z=5$.

In the \ninja runs, we look for the convergence of the BHMF for both the central (solid green)
and satellite (dashed green) populations.
Comparing the L140 series of the \ninja runs, we find that the BHMF of the intermediate run (L140\_N896) is converged for $\log_{10}(\Mbh/\msun) \gtrsim 7.5$  for the central and satellite populations.
The BHMF in the lowest resolution run (L140\_N700)  converges for   $\log_{10}(\Mbh/\msun) \gtrsim 8.5$. Comparing the fiducial (L140\_N1008) with the high-resolution run (L50\_N640), we find that the
BHMF of the fiducial run converges for $\log_{10}(\Mbh/\msun) \gtrsim 6.5$, which is about one order of magnitude larger than the seed mass. 
The BHMF of the high resolution run does not extend beyond  $\log_{10}(\Mbh/\msun) \geq 9.2$ and is suppressed at  $\log_{10}(\Mbh/\msun) \geq 8.5 $ in comparison to the L140 series which
is due to finite volume effects. We see that the BHMF of the L140 series is suppressed for $\log_{10}(\Mbh/\msun) \geq 8.5$ with respect to \astrid. This may be due to finite volume effects or due to the differences in the feedback prescriptions between both 
the simulations. 
When we consider all MBHs in the high, medium, and fiducial \ninja runs, by including the wandering black holes, we find an excellent match between \ninja and \astrid at masses
$\log_{10}(\Mbh/\msun) \leq  7.0$. This suggests that black hole growth in \ninja is not affected due to the relatively poorer resolution in \ninja. In  section~\ref{subsec_timedelay},
we discuss a post-processing model of delaying the time of merger, due to the limited spatial resolution of cosmological simulations. This requires properties of
the host galaxy. For our black hole resolution limit,   $\log_{10}(\Mbh/\msun) \geq 6.5$, the host galaxy in the fiducial run is resolved on average with 320 star particles, and at $2\sigma$ is resolved with 115
star particles. In the rest of the paper, we only consider MBHs with    $\log_{10}(\Mbh/\msun) \geq 6.5$. We discard less than a percent of the total number of MBH mergers due
to the absence of a corresponding host galaxy in these low-mass MBHs. 
We will show that the GW signal is detectable in LISA only for a total binary mass of the black hole $\log_{10}(M_{\rm bin}/\msun) \leq 8.4$, hence differences between \astrid and \ninja at large masses should not lead to any loss of detectable events.

\ninja shows an excellent agreement with the observations of \cite{2014ApJ...786..104U} and \cite{Li_2011}.
\illustris matches the best with the observations of \cite{2013CQGra..30x4001S}. \astrid and \tngt also match these observations for  $\log_{10}(\Mbh/\msun) \geq 8.0$.
However, there is a lack of convergence between \tngh and \tngt, which may improve if one removes wandering black holes and includes a resolution criterion for the host galaxy. Similarly we believe that if wandering black holes were removed 
in \astrid it would bring it closer to the observations of \cite{2014ApJ...786..104U} and \cite{Li_2011} at intermediate masses.

\subsection{Black hole and host galaxy merger trees}

\begin{figure}[t]
  \centering
  \includegraphics[width=1\linewidth]{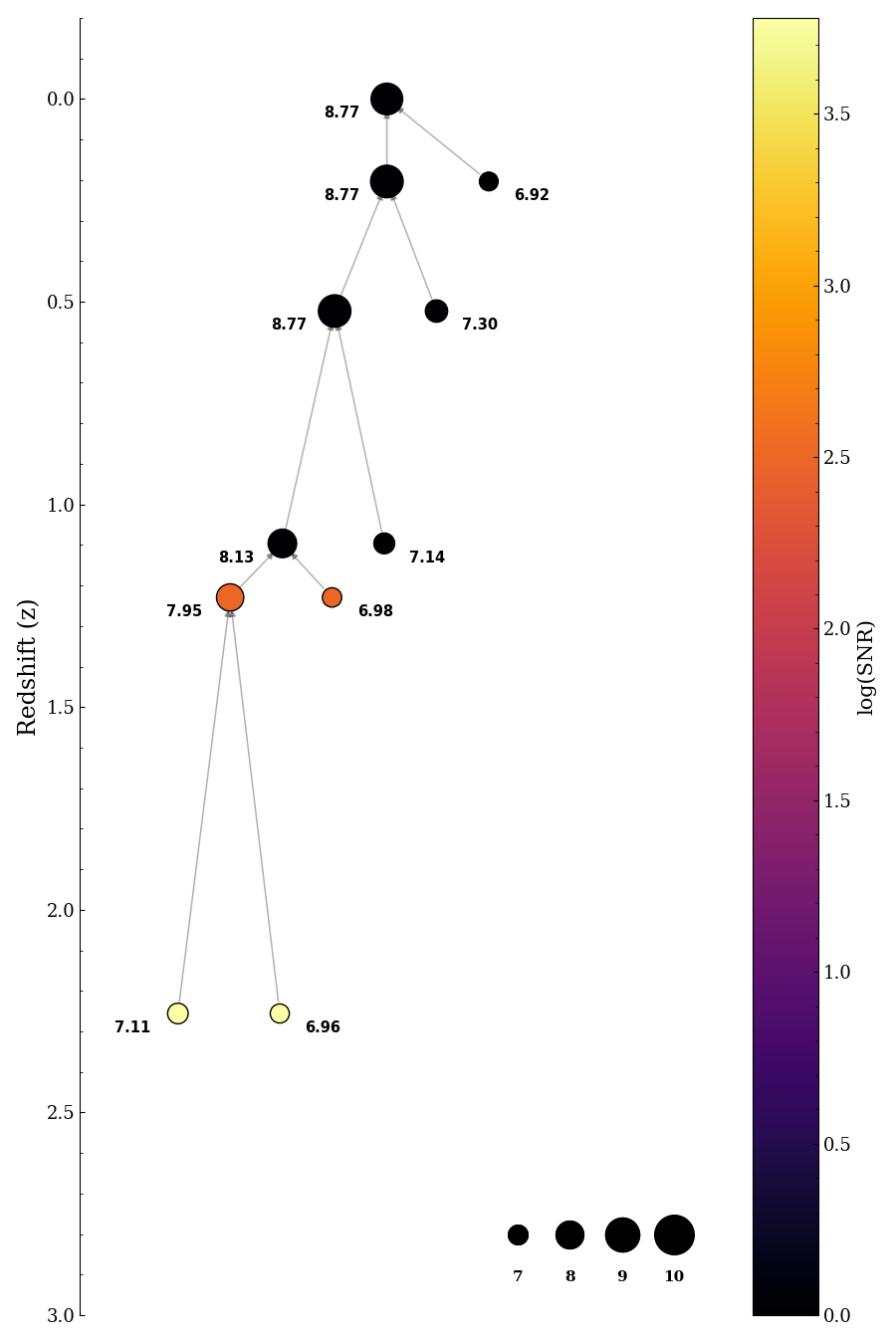}
  \caption{An example of a merger tree tracing the evolution of a progenitor black hole through time. The y-axis represents redshift ($z$), with
    cosmic time increasing downwards. Each circle represents a black hole, with its size proportional to the logarithm of the mass of the
    black hole $\log_{10}(\Mbh/\msun)$. The color of each black hole at a given redshift is determined by the logarithm of the
    signal-to-noise ratio (log(SNR)) as detected by LISA, with the color bar on the right indicating the scale.
    The connecting lines illustrate the hierarchical merging of black holes over cosmic time.}
  \label{fig:merger_tree_ex}
\end{figure}

Unlike other particles, \mpgadget stores the properties of all black holes that are active, including merger information of the MBH pair and their local environment.
In order to track host galaxy properties of merging MBHs, which are required for modeling time delays, we store about 170 snapshots (100 in logarithmic intervals of the scale factor from
$z=10$ to $z=0$, interspersed
with 70 more at unit, half, quarter, and a tenth of redshift intervals). This allows us to look at the light curves of each individual MBH, before, during, and after a merger.
It also allows us to look at evolution in the properties of the host galaxies. 

In figure~\ref{fig:merger_tree_ex} we illustrate the merger tree of a MBH of mass $\log_{10}(\Mbh/\msun) = 8.77$ at $z=0$. The history of the progenitor (the most massive
of two merging MBHs) is traced back in time (vertical axis). The radius of the circle, representing the MBH, is proportional to
the logarithm of the mass of the MBH in solar masses. The color represents the signal-to-noise ($\log_{10}({\rm SNR})$) as detected in LISA during a merger.
This example shows that the GW signal from very massive black holes do not fall in the detection threshold of LISA for recent mergers, but  earlier mergers are detectable
when the progenitor was less massive. 

\begin{figure}
\includegraphics[width=\columnwidth]{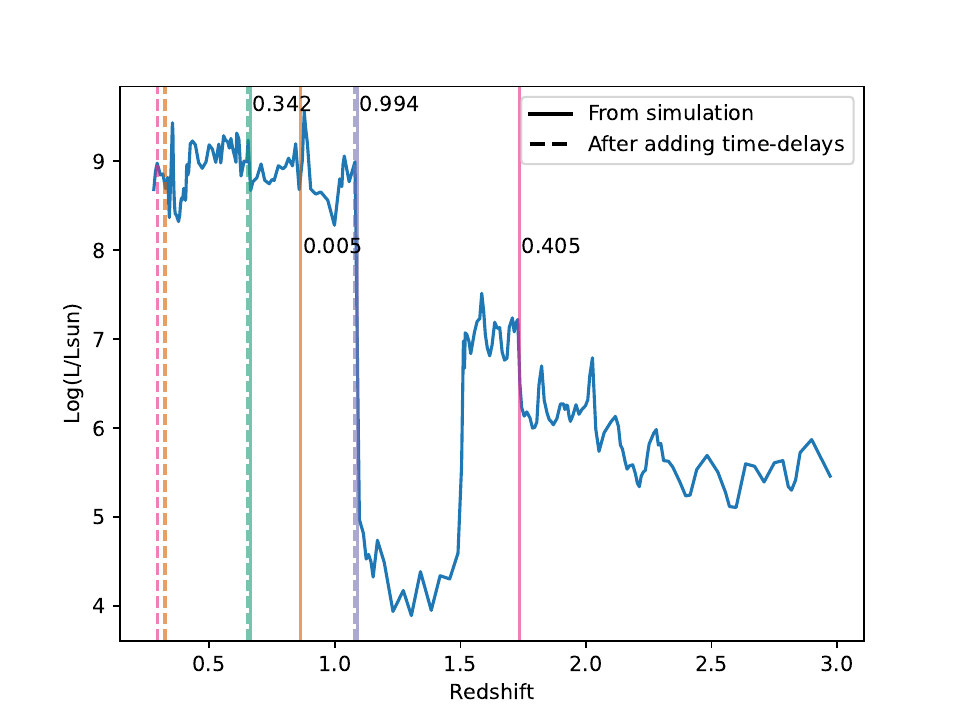}
\caption{ Bolometric luminosity as a function of redshift plotted by tracking an ID of a black hole from the merger tree.
  Each vertical line of distinct color marks a MBBH merger. The solid line corresponds to the merger where the MBBHs are considered merged by the simulation. The number beside each merger denotes the mass ratio of the merger.
  Dashed line corresponds to the merger redshift after adding time delays due to dynamical friction.}
\label{fig:accretion_rate_ex}
\end{figure}

We limit our analysis to $z \leq 2.5$ since the number of mergers is extremely small at higher redshifts, as we will see in section~\ref{sec_results}.
In figure~\ref{fig:accretion_rate_ex} we look at the mass accretion history of a MBH at $z=0$, described by its bolometric luminosity in solar units.
The solid vertical line with a number beside it denotes a merging event as identified in the simulation and its mass ratio. The dashed vertical line
denotes the merger redshift after adding time delays. Mergers are typically associated with bursts in the luminosity (or accretion rate), which can increase 
by as much as four orders in magnitude in this particular example. This is because fresh gas associated with the host galaxy of the black hole can efficiently
reach the local, dense environment of the merging black hole and get accreted onto it more efficiently. 

\begin{figure*}
\centering
\includegraphics[width = 1\textwidth]{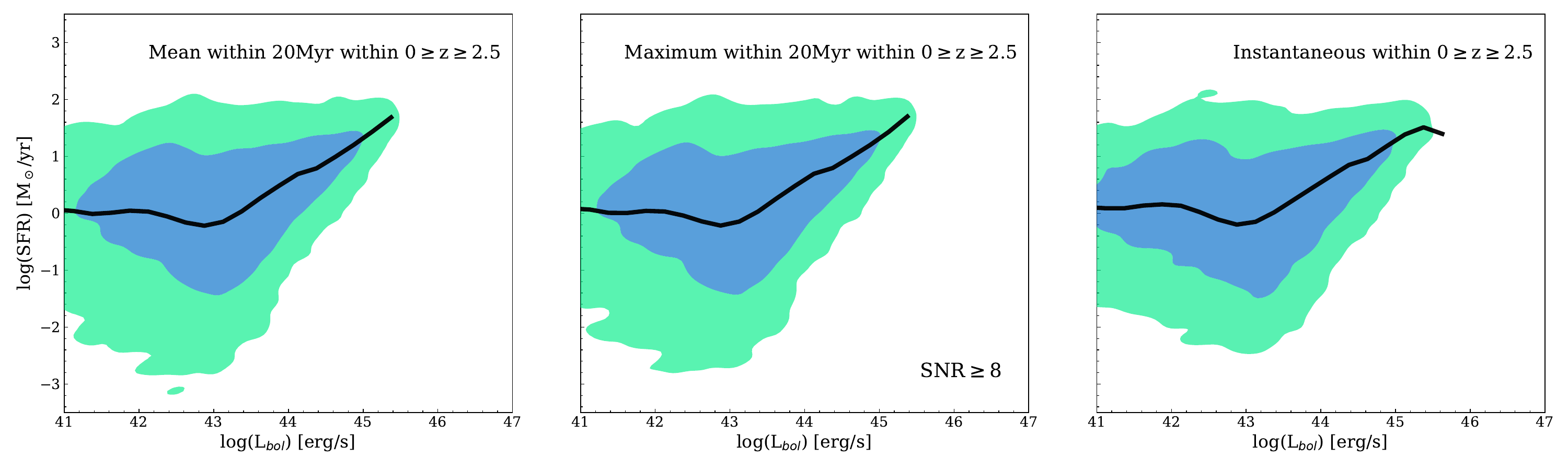}
\caption{A comparison of the different methods (panels left to right) used to calculate the SFR of galaxies hosting merging black holes systems with
  bolometric luminosity, $L_{\rm bol}$, with a GW SNR greater than 8 for the redshifts $0 \leq z \leq 2.5$.
  \textit{Left}: SFR and $L_{\rm bol}$ are calculated using their mean values within 20 Myr time merger.
  \textit{Middle}: SFR and $L_{\rm bol}$ based on their maximum values within  20 Myr of the merger.
  \textit{Right}: SFR and $L_{\rm bol}$ based on the closest snapshot data to the merger.
  The solid black line in each panel represents the median relation. The contours represent the 68\%, 95\% of the enclosed population.}
\label{fig_sfr-lbol_all}
\end{figure*}

In figure~\ref{fig_sfr-lbol_all} we look at the star formation rate (SFR) of galaxies hosting merging black hole systems with bolometric luminosity ($L_{\rm bol}$) resulting in a GW with SNR greater than 8 for redshifts $0 \leq z \leq 2.5$. The panels compare different estimates of these quantities. As discussed earlier, MBH properties are stored at every active timestep, which is on average 0.14 Myr for the most active black hole. Galaxy and halo properties are stored at only predetermined epochs which on average are spaced every $\sim$50 Myr. MBH mergers can happen in between snapshots; therefore, the SFR-$L_{\rm bol}$ scaling relation can be a sensitive function of the exact time when each of the properties are extracted, which in general are not synchronised. We have seen in figure~\ref{fig:accretion_rate_ex} that the bolometric luminosity of the MBH can change by orders of magnitude in a short time around a merger. Galaxy properties, on the other hand, do not change so rapidly in such short periods of time. 

In order to synchronize both the SFR and $L_{\rm bol}$ around a merger, we implement a windowed Gaussian smoothing interpolation scheme. For each merger event, we define a temporal window and interpolate both SFR and $L_{\rm bol}$ onto a common lookback time with 1 Myr resolution using a weighted averaging approach. Within each 1 Myr interval, data points are selected from a sliding window and weighted by a Gaussian kernel, where the width of the sliding window is $\sim$25 Myr for galaxy properties and $\sim$10 Myr for MBH properties. These widths reflect the characteristic variability timescales of the respective physical processes. This approach effectively synchronizes the time series while preserving the underlying physical correlations between star formation and MBH activity during merger events.

In the left panel of figure~\ref{fig_sfr-lbol_all}, the SFR and $L_{\rm bol}$ are calculated using their mean values within 20 Myr of merger.
In the middle panel, the SFR and $L_{\rm bol}$ are based on their  maximum values within 20 Myr of the merger, and in the right panel
the SFR and $L_{\rm bol}$ are extracted from the closest snapshot to the time of merger.
The solid black line in each panel represents the median relation.
We find that the left and middle panels are comparable.
The $1\sigma$ contour is more compact in the left and middle panels compared to the right panel; however, the median
relation is comparable across the panels. This suggests that the median relation is not very sensitive
to which scheme, interpolation (left and middle panels) or instantaneous (right panel), is used to evaluate the
properties. However, the SFR-$L_{\rm bol}$ relation has less scatter in the interpolation scheme.

\subsection{Extraction of time-delay}
\label{subsec_timedelay}

Dissipation of energy by GW radiation is dominant when the separation between the MBHs is of the order of
milliparsecs and makes them merge within Hubble time \citep{2014SSRv..183..189C}. However, the simulation numerically merges the MBHs at much larger distances than milliparsec.
Therefore, the simulation data has to be post-processed in order to make statements about the merging time of MBH binaries.
Several methods for incorporating the merging timescale in numerical simulations have been suggested
\citep[][]{2017MNRAS.464.3131K, 2019ApJ...879..110K, 2020MNRAS.491.2301K, 2021MNRAS.501.2531S}.

\ninja already accounts for the dynamical friction time scale above the resolution limit. This dynamical friction model is an alternative to the repositioning model, which constantly repositions the BH position to the potential minimum and was used in earlier simulations \citep[see][]{2022MNRAS.510..531C}. Including the dynamical friction model leads to significant delays in the in-simulation merger time of the MBHs as compared to the repositioning model. However, the dynamical friction continues to act on the BHs even on the scale of $\mathcal{O}(10)-\mathcal{O}(100)$ pc, which is beyond the resolution limit of the simulation. This effect of dynamical friction is accounted for in the post-processing step of the simulation.

In this work, we assume the dynamical friction and stellar hardening as the principal mechanisms for the orbital evolution of MBHs
after the merging of galaxies. Initially, MBHs are tightly bound to their individual stellar bulges so the bulge and MBH together act as a single object.
These parent bulges sink towards each other until they reach $\mathcal{O}(1kpc)$ and the bulges merge. Further, as long as the mass of the stars and
the gas between the MBHs is much larger than the total mass of MBHs, they are continually dragged towards each other by the
dynamical friction exerted by the background stars and gas of the new galaxy. \citep[see][]{2022LRR....25....3B} 
We assume that the MBHs follow a circular orbit, and the stars around them follow an isothermal sphere, then the time delay due to
dynamical friction is given by \citep[see][]{2008gady.book.....B, 2019ApJ...879..110K, 2020MNRAS.498.2219V},

\begin{multline}    
\label{eq:dynamical_friction} 
t_{\rm df} = 0.672 \left( \frac{a}{4 \rm kpc}\right)^{2} \left(\frac{\sigma}{100 {\rm km/s}} \right) \left(\frac{10^{8} \msun}{M_{2}}  \right) \\
\times \frac{1}{\log_{10}(1 + M_{\rm gal}/M_{2})}
\end{multline}
where $M_{\rm gal}$ and $\sigma$ are the total stellar mass of the galaxy and 
the central stellar velocity dispersion approximated as $(0.25GM_{\rm gal}/R_{\rm eff})^{1/2}$, where $R_{\rm eff}$ is the half mass radius of the galaxy.
$M_{2}$ is the mass of the smaller black hole.
$a$ is the distance between $M_{2}$ and galaxy centre, calculated in the simulation.
In the above expression, we have multiplied by a factor of $0.3$ to account for orbits being non-circular \citep{2003MNRAS.341..434T}. 

After the time delay due to dynamical friction has elapsed, the black holes are close enough that their total mass is significantly larger
than the mass of the gas and stars within their orbit. The main dynamical driver considered in this work is scattering by individual stars.
We calculate the stellar hardening timescale as given by \citep{2015MNRAS.454L..66S, 2020MNRAS.498.2219V}.
\begin{equation}
t_{\rm bin} = 15.18 \rm{Gy} \left(\frac{\sigma_{\rm inf}}{\rm km/s}\right)\left(\frac{\rho_{\rm inf}}{\msun {\rm pc}^{-3}}\right)^{-1} \left(\frac{a_{\rm gw}}{10^{-3}pc}\right)^{-1}
\end{equation}
where $\sigma_{\rm inf}$ and $\rho_{\rm inf}$ are the velocity dispersion and stellar density at the sphere of influence,
defined as the sphere containing twice the binary mass in stars. $a_{\rm gw}$, which is the separation at which the binary spends most time, is given by

\begin{multline}    
a_{\rm gw} = 2.64\ \times 10^{-2} \rm{pc}  \\
\left[\frac{\sigma_{\rm inf}}{\rm km/s} \left(\frac{\rho_{\rm inf}}{\msun {\rm pc}^{-3}}\right)^{-1} \frac{15}{H}\left(\frac{M_1M_2(M_1 + M_2)}{2\times 10^{24} \msun^3}\right)\right]^{1/5}
\end{multline}
We use $H=15$ as the reference value. The parameters $\sigma_{\rm inf}$ and $\rho_{\rm inf}$ are derived by assuming a power-law density
profile inside $R_{\rm eff}$ where $\rho_{\rm inf} \propto r^{-2}$ with a total mass within $R_{\rm eff}$ equal to half the galaxy's stellar mass.
Using this power-law profile and the black hole masses $M_{1} + M_{2} = M_{\rm BH}$, we can compute the radius $r_{\rm inf}$ containing
twice the binary mass in stars, the density at that radius, $\rho_{\rm inf}$, and velocity dispersion as $\sigma_{\rm inf}$
as $(GM_{BH}/r_{\rm inf})^{1/2}$ \citep[see][]{2020MNRAS.498.2219V}.

\subsection{Gravitational Wave detection by LISA}
Mergers of MBHBs can produce GWs in the milli- and nano- frequency range, which can be detected by LISA and Pulsar Timing Array (PTA), respectively. In this paper, we focus only on the LISA detector and the properties of the MBH population that it will detect.

LISA \citep[][]{2001AcAau..48..549E, 2017arXiv170200786A, 2024arXiv240207571C} 
is a space-based GW observatory which is planned to be launched in the next decade. This observatory will consist of
three satellites forming a triangular constellation in a heliocentric orbit. In each satellite, freely falling test masses will
be placed which will serve as inertial reference points that define the end of a measurement arm, with each arm
being millions of kilometers in length. 
In this section, we describe the construction of the LISA sensitivity curve and the calculation of the signal-to-noise ratio.

The sensitivity curve or the power spectral density (PSD) is defined as the Fourier transform of the autocorrelation
function of noise in the data. Intuitively, the height of the signal above the PSD will determine the loudness of the signal
above the background noise. We use \citep{2019CQGra..36j5011R} to construct the PSD and calculate the SNR.

The LISA PSD can be well approximated by this expression:

\begin{multline}
S_{\rm n}(f) = \frac{10}{3L^{2}}\left(P_{\rm OMS}(f) + \frac{4P_{\rm acc}(f)}{(2\pi f)^{4}}\right)\times
\\
\left(1 + \frac{6}{10}\left(\frac{f}{f_{*}}\right)^{2}\right) + S_{\rm c}(f)
\end{multline}

where $L = 10$Gm , $f_{*} = 19.09$mHz, $P_{\rm OMS}(f)$ is the single-link optical metrology noise, $P_{\rm acc}(f)$ is the single test mass acceleration noise. The exact values of these quantities can be found in the
reference \citep{2001PhRvD..65b2004C,2019CQGra..36j5011R}. $S_{\rm c}(f)$ is the galactic confusion noise because of the unresolved Galactic binaries. The estimations of the confusion noise for LISA is given in reference \citep{2017JPhCS.840a2024C, 2019CQGra..36j5011R} and can be well approximated by the function:

\begin{multline}
S_{\rm c}(f) = Af^{-7/3}\exp{\left(-f^{\alpha} + \beta f \sin{(\kappa f)}\right)}  \\ 
\times \left(1 + \tanh{\gamma (f_{k} - f)} \right)  \rm Hz^{-1}
\end{multline}

As the Galactic confusion noise reduces with observation time, the value of the parameters in the above fitting function- $\alpha, \beta , \gamma, \kappa, f_{k}$
also changes with the observation time. In this work, we choose the observation time of $4$ years. The corresponding values of the confusion noise parameters are given by $\alpha = 0.138$, $\beta = -221 $, $\gamma = 1680$, $\kappa = 521$, $f_{k} = 0.00113$.

LISA will detect GWs from the MBHBs as well as extreme mass ratio binaries. In this analysis,
we have simplified the treatment by considering binaries to be quasicircular and non-spinning systems, dominated by the leading order $22$ mode. For this system of binaries, the waveforms for $+$ and $\times$ polarization is given by:

\begin{equation}
\tilde{h}_{+}(f) = A(f)\frac{(1 + \cos^{2}(\iota))}{2}e^{i \Psi(f)}
\end{equation}
\begin{equation}
\tilde{h}_{\times}(f) = i A(f) \cos(\iota) e^{i \Psi(f)}
\end{equation}

where $A(f)$ and $\Psi(f)$ are the amplitude and phase of GW respectively and $\iota$ is the inclination angle. 

In this work, we use the IMRPhenomA model, which is a phenomenological Inspiral-Merger-Ringdown (IMR) model for waveform modelling \citep{2007CQGra..24S.689A}.
The amplitude $A(f)$ for each phase of the binary-inspiral, merger, and ringdown is given by,

\begin{multline}
A(f) \equiv \sqrt{\frac{5}{24}} \frac{(G\mathcal{M}/c^3)^{5/6} f^{-7/6}_{0}}{\pi^{2/3}(D_L/c)} \times \\
\begin{cases}
\left(\frac{f}{f_0}\right)^{-7/6} & \text{if } f < f_0 \\
\left(\frac{f}{f_0}\right)^{-2/3} & \text{if } f_0 \leq f < f_1 \\
w \mathcal{L}(f, f_1, f_2) & \text{if } f_1 \leq f < f_3,
\end{cases}
\end{multline}

where

\begin{equation}
\hspace{1.5cm}
f_{\rm k} \equiv \frac{a_k \eta^2 + b_k \eta + c_k}{\pi (GM/c^3)},
\end{equation}

\begin{equation}
\hspace{1.5cm}
\mathcal{L}(f, f_1, f_2) \equiv \left(\frac{1}{2\pi}\right) \frac{f_2}{(f - f_1)^2 + f_2^2/4},
\end{equation}

\begin{equation}
\hspace{1.5cm}
w \equiv \frac{\pi f_{2}}{2}\left(\frac{f_{0}}{f_{1}} \right)^{2/3}
\end{equation}

In the above equations $M = m_{1} + m_{2}$ is the total mass of the systems, $\eta = m_{1}m_{2}/M^{2}$ is the symmetric mass ratio, $\mathcal{M} = (m_{1}m_{2})^{3/5}/M^{1/5}$
and $D_{L}$ is the luminosity distance to the merger. The coefficients for the transition frequencies  $f_{\rm k}$ are given by the
Table \ref{transition_freq_table}. Here, $f_{0}$ is the merger frequency, $f_{1}$ is the ringdown frequency, $f_{2}$ is the decay width of the ringdown, and $f_{3}$ is the cutoff frequency, i.e., the frequency which marks the end of GW emission from CBCs.
\begin{table}[]
\centering
\begin{tabular}{|c|c|c|c|}
\hline
& $a_{k}$ & $b_{k}$ & $c_{k}$ \\ 
\hline
$f_0$ & $2.9740 \times 10^{-1}$ & $4.4810 \times 10^{-2}$ & $9.5560 \times 10^{-2}$ \\
$f_1$ & $5.9411 \times 10^{-1}$ & $8.9794 \times 10^{-2}$ & $1.9111 \times 10^{-1}$ \\
$f_2$ & $5.0801 \times 10^{-1}$ & $7.7515 \times 10^{-2}$ & $2.2369 \times 10^{-2}$ \\
$f_3$ & $8.4845 \times 10^{-1}$ & $1.2848 \times 10^{-1}$ & $2.7299 \times 10^{-1}$ \\
\hline
\end{tabular}
\caption{Coefficients $a_{k}, b_{k}, c_{k} $ for different transition frequencies}
\label{transition_freq_table}
\end{table}

To calculate the signal-to-noise ratio (SNR) of a binary, we first need to compute GW amplitude $\tilde{h}(f)$ in the detector, which is given by:

\begin{equation}
\tilde{h}(f) = F_{+}(\theta,\phi,\psi,f)\tilde{h}_{+}(f) + F_{\times}(\theta,\phi,\psi,f)\tilde{h}_{\times}(f)
\end{equation}
where $F_{+}(\theta,\phi,\psi,f)$ and $F_{\times}(\theta,\phi,\psi,f)$ are the antenna pattern functions, which depend on sky location $(\theta,\phi)$
and the polarization angle $\psi$ of the source. The optimal SNR for a given signal with GW amplitude $\tilde{h}(f)$ is given by:
\begin{equation}
\rho^{2} = 4\int \frac{|\tilde{h}(f)|^{2}}{S_n(f)} df
\end{equation}

We average the SNR over the sky-location and polarization angle, and assume the inclination angle to be optimally oriented. The optimal SNR is given by:
\begin{equation}
\label{avg_snr_eq}
\langle \rho^{2} \rangle = 4 \int_{f_{\rm low}}^{f_{\rm high}} \frac{A^{2}(f)}{S_{\rm n}(f)} df
\end{equation}

For all computations, we consider the contribution of each phase i.e., inspiral, merger and ringdown to the SNR only if that
particular phase lies in the LISA band with $f_{\rm low} = 10^{-4}$ Hz. For e.g. 

\begin{equation*}
\hspace{1cm}
\text{if} \hspace{0.2cm} f_{0} > f_{\rm low}: \hspace{0.5cm}\rho^{2} = \rho^{2}_{\rm insp} + \rho^{2}_{\rm merg} + \rho^{2}_{\rm ring}   
\end{equation*}

\begin{equation*}
\hspace{1cm}
\text{if} \hspace{0.2cm} f_{0} < f_{\rm low} \hspace{0.2cm} \text{and}   \hspace{0.2cm} f_{1} > f_{\rm low} : \hspace{0.5cm}\rho^{2} = \rho^{2}_{\rm merg} + \rho^{2}_{\rm ring}   
\end{equation*}

\begin{equation*}
\hspace{1cm}
\text{if} \hspace{0.2cm} f_{0} < f_{\rm low} \hspace{0.2cm} \text{and}   \hspace{0.2cm} f_{1} < f_{\rm low} : \hspace{0.5cm}\rho^{2} = \rho^{2}_{\rm ring}   
\end{equation*}

\begin{equation*}
\hspace{1cm}
\text{if} \hspace{0.2cm} f_{3} < f_{\rm low} : \hspace{0.5cm}\rho^{2} = 0   
\end{equation*}

$f_{\rm high}$ is the GW frequency of the binaries after 4 years of observing time. 

\begin{figure*}[t]
\centering 
\includegraphics[scale = 0.55]{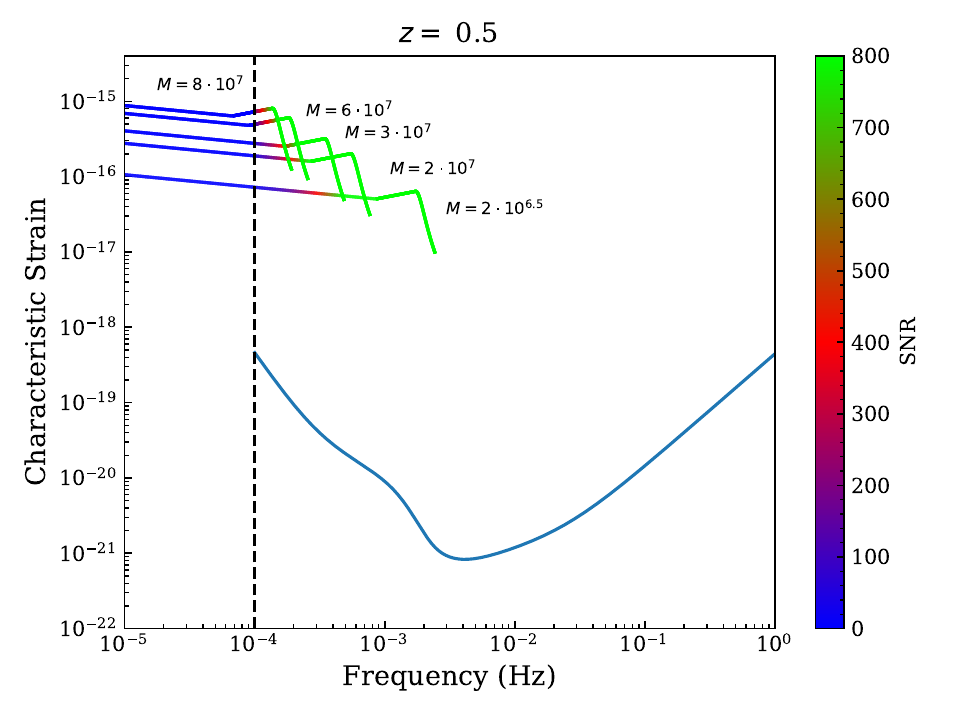}
\includegraphics[scale = 0.55
]{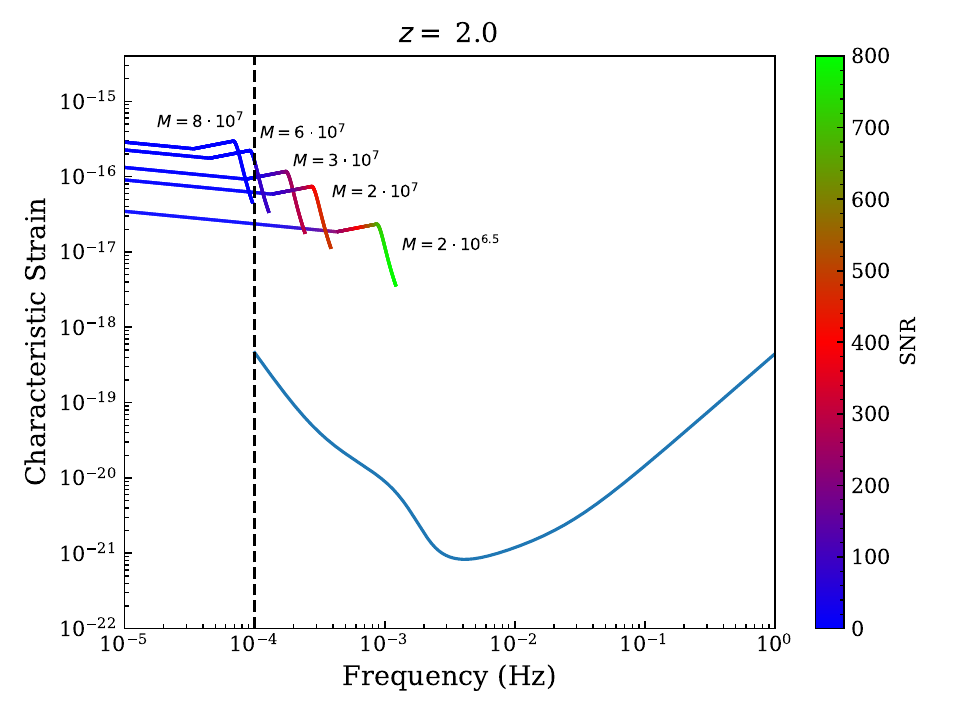}
\caption{Characteristic strain plot for GW signal for different total masses at redshift $0.5$ and $2.0$. The color indicates the SNR of the signal.
  The dashed line indicates the lower limit of frequency for LISA.}
\label{fig:char_strains}
\end{figure*}

To compute SNR, we require masses of the binaries and the luminosity distance at which they merged. We pick out the binaries that the simulation considers to be merged,
from which we get the masses of the merged binaries and the scale factor at which they are merged. We then convert the masses to the detector
frame masses by $M_{z} = M(1 + z)$, and we  
convert the redshift to the luminosity distance.
Because of the resolution of the simulation, we put a lower mass cut of $10^{6.5}\msun$ on the BH mass. We then calculate the SNR using Equation \ref{avg_snr_eq}.

Fig. \ref{fig:char_strains} illustrates the characteristic strain and SNR for GW signals emitted by binary black holes (BBHs) at redshifts of $z =$ $0.5$ and $ 2.0$, across a range of total masses. The figure plots frequency on the x-axis and
characteristic strain on the y-axis, featuring multiple curves, each corresponding to a different BBH mass. These curves trace the GW signal's evolution through three
distinct phases: the inspiral phase, depicted as a nearly horizontal line on a log-log scale, reflecting the gradual orbital decay of the binary;
the merger phase, where the strain reaches its peak as the black holes coalesce, and the ringdown phase, characterized by a subsequent
decrease in strain as the merged black hole stabilizes. Colors along each curve indicate the SNR for these phases, quantifying the signal's detectability
relative to the noise of the LISA detector. A dashed vertical line marks \( f_{\text{low}} \), LISA's lower frequency limit (0.1 mHz), which is pivotal since
the SNR is computed by integrating the signal from \( f_{\text{low}} \) to the end of the ringdown phase. For BBHs with higher total masses, the GW signal shifts
to lower frequencies, often falling below \( f_{\text{low}} \), resulting in reduced SNRs, whereas lower-mass systems produce signals
within LISA's sensitive range (0.1 mHz to 1 Hz), yielding higher SNRs. 
This figure thus highlights the critical dependence of GW detectability on the interplay
between BBH mass, signal frequency evolution, and LISA's sensitivity band, emphasizing the difficulties in observing
very high-mass binaries and delineating the optimal mass range for maximizing SNR. 

Although the MBHs in general have a non-zero value of spins, and we have accurate waveforms like IMRPhenomP \citep{2014GReGr..46.1767H,2020PhRvR...2d3096P}
to model them, for our calculation of GW strain, we are ignoring the spin of the black holes. We compared our analysis by adding amplitude
corrections because of spins, as mentioned in \citep[][]{2016PhRvD..93d4007K} to our non-spin calculation, we find that for the case when both
MBHs have maximum spins; the fractional increase in SNR is only about $\sim 1$ percent.

\section{Results and Discussion} 
\label{sec_results}
\subsection{Detectable population by LISA}

\begin{figure*}[p]
\centering
\includegraphics[scale = 0.5]{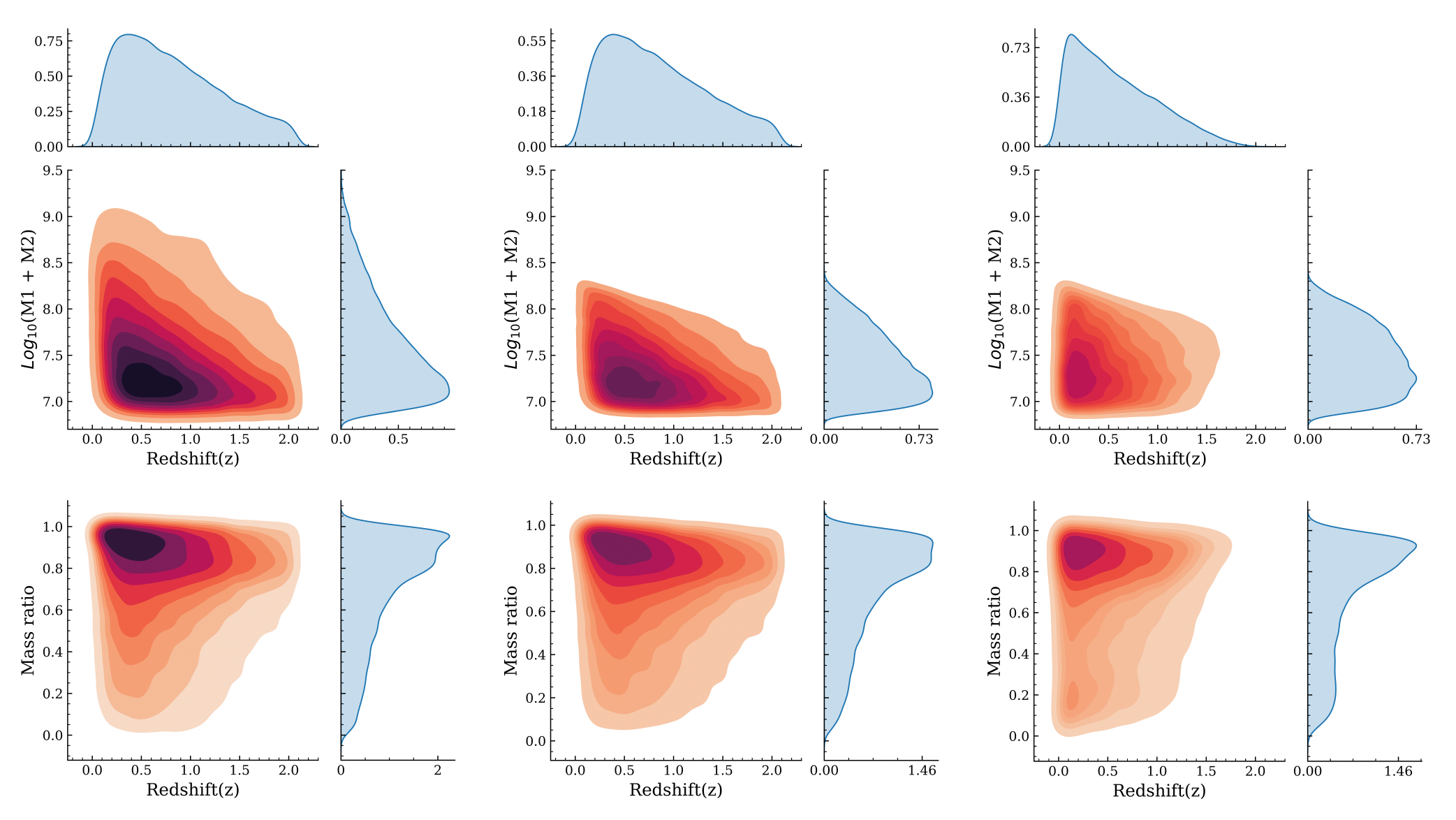}
\caption{Distribution of the Total mass (\textit{top}) and Mass ratio (\textit{bottom}) of BBH mergers versus the redshift is plotted
  along with the corresponding marginalized distributions. 
  \textit{Left}: The distribution of total mass and mass ratio of BBH merger from the simulation. \textit{Middle}: The distribution of total mass and the
  mass ratio of the BBH merger, which are detectable in LISA, and \textit{Right}: The distribution of total mass and mass ratios detected by LISA after adding time delays due to dynamical friction and stellar hardening. }
\label{fig:LISA_detectable_with_delays}
\end{figure*}

\begin{figure*}[p]
\centering
\includegraphics[scale = 0.5]{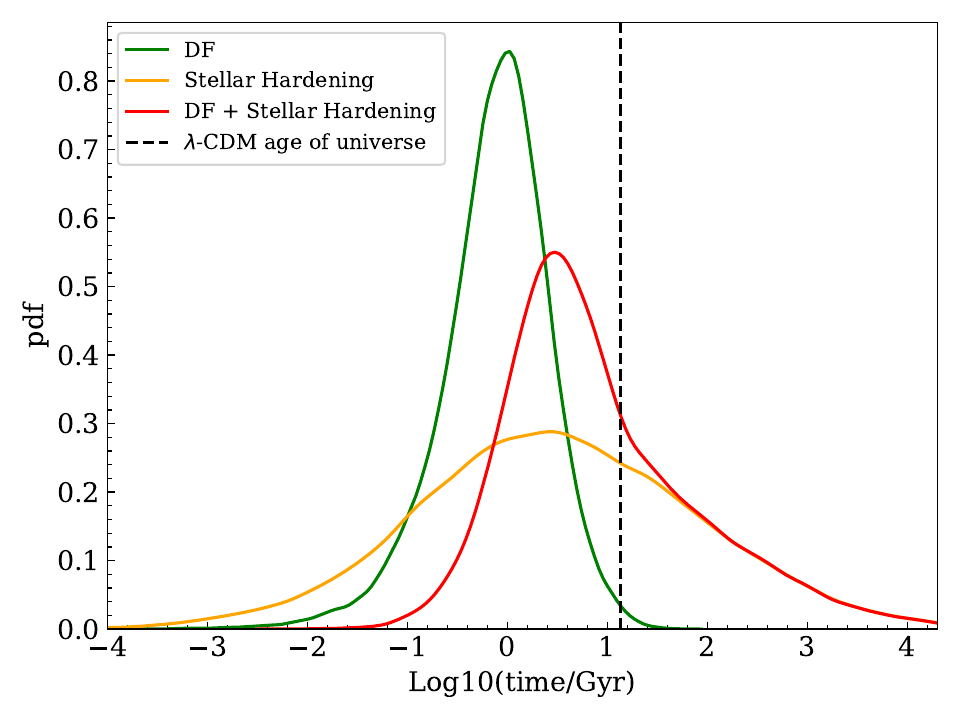}
\includegraphics[scale = 0.5]{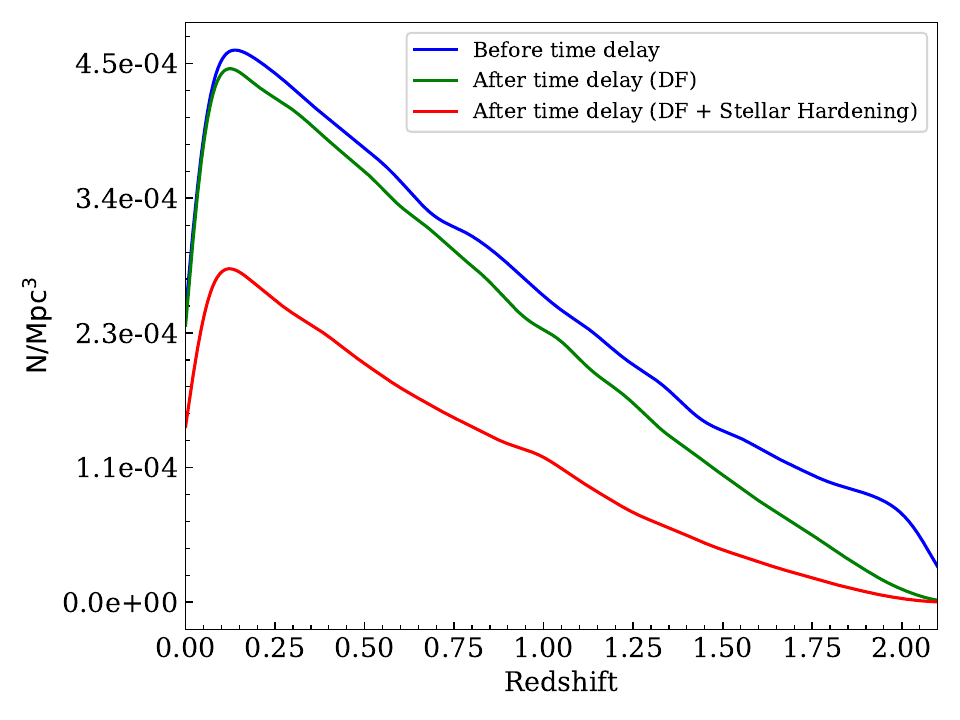}
\caption{\textit{Left}: The time delay distribution due to Dynamical friction (\textit{green}), Stellar Hardening (\textit{orange})
  and by both (\textit{red}). \textit{Right}: Redshift distributions till $z=0$ of BBH mergers before and after the time delay.}
\label{fig:Time_delay_distribution}
\end{figure*}

We first check the detectability of the MBHBs in the LISA detector based on SNR calculations shown in the previous section. 

For each MBH merger in the simulation, we get the mass of BHs and the redshift at which they have merged. We convert the redshift to the luminosity distance and then calculate the SNR. However, this is the SNR of the MBHBs without considering the time delays because of dynamical friction and stellar hardening. We consider these time delays in the post-processing step by first converting the redshift of the merger to look-back time and then subtracting from it $t_{\rm df} + t_{\rm bin}$ 

In Fig. \ref{fig:LISA_detectable_with_delays}, we show the distributions of the total mass and the mass ratio of the MBHBs as functions of the redshift at which the merger occurs. For the purpose of this analysis, we have chosen the detectability criterion of MBHBs by LISA if the SNR exceeds 8. The SNR is determined by the amplitude and frequency
content of the GW signal within LISA's sensitive frequency band, which spans approximately $10^{-4}$ to $10^{-1}$ Hz. Further, our calculation of SNR also includes the SNR from the merger and the ringdown part of the GW signal, along with the SNR from the inspiral. This boosts the detectability of the merger and also the analysis is more rigorous than the ones that only include the inspiral part of the GW signal.

In Fig.\ref{fig:LISA_detectable_with_delays}, before adding the time delays, we can compare the total mass of all the MBHBs from the simulation to the detectable ones in LISA. It is evident that MBHBs possessing a total mass greater than approximately $10^{8.4}$ solar masses ($\msun$)
fall outside the detection capabilities of LISA. Upon comparing the distribution of mass ratios for the detectable MBHBs, we observe that it does not exhibit any
significant deviations from that of the total population of MBHBs from the simulation. Moreover, the redshift distribution of the detectable
binaries mirrors that of all simulated binaries. This suggests that the primary factor limiting the detectability (SNR $\geq$ 8) of GWs by LISA is not the distance to
the binary system, but rather the total mass of the MBHBs. The underlying reason for this phenomenon is that binaries with greater total mass undergo mergers
at lower frequencies. Specifically, the characteristic GW frequency at merger, often associated with the frequency at the innermost stable circular orbit (ISCO), scales inversely with the total mass of the binary. Consequently, for MBH binaries with total masses exceeding $10^{8.4} \msun$, this frequency lies below LISA's lower frequency threshold, making them undetectable. In such cases, these massive mergers do not contribute to the SNR within LISA's operational bandwidth.

Upon adding the time delays due to dynamical friction and stellar hardening for all the population of MBHBs, we observe a change in the redshift distribution of the population detectable by LISA.
As seen in Fig. \ref{fig:LISA_detectable_with_delays}, the peak in the redshift distribution of the detectable population without adding time delays is at $z=0.5$. After taking into account the time delays, the peak of the redshift distribution shifts to $z=0.1$. This change occurs because incorporating dynamical friction and stellar hardening causes the time delays to be $\mathcal{O}(1)$ Gyr, pushing the merger event to later cosmic epochs, which correspond to lower redshift values. 

\begin{figure}[t]
\centering
\includegraphics[width=1\linewidth]{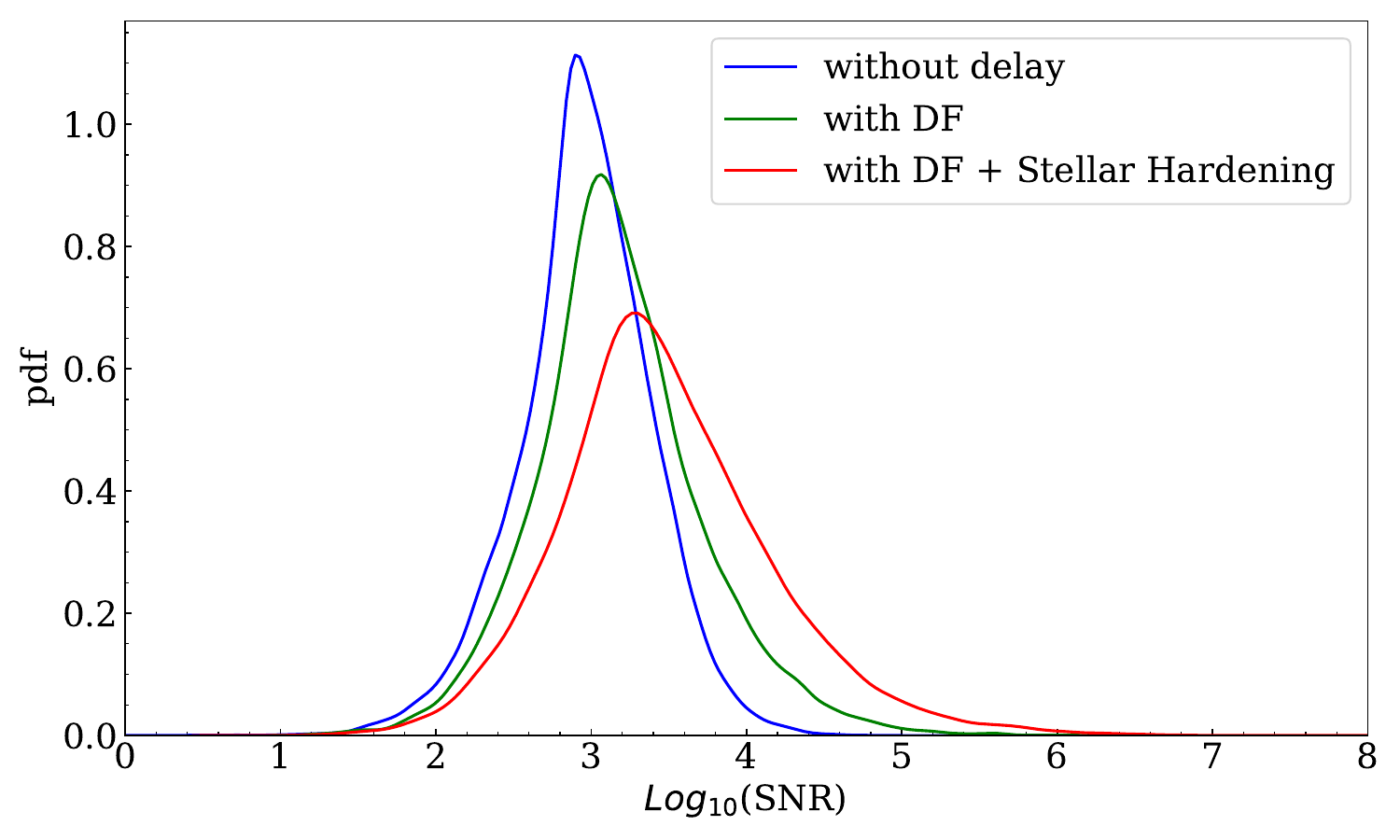}
\caption{Histogram of SNRs of binaries with different postprocessing scenarios- \textit{Blue} showing without adding time delays. \textit{Green} showing after adding dynamical friction time delay and \textit{Red} after adding time delays due to dynamical friction.}
\label{fig:SNR_histogram}
\end{figure}

The distribution of time delays because of dynamical friction and stellar hardening is illustrated in Fig.\ref{fig:Time_delay_distribution}. It shows a pronounced peak at approximately $1$ Gyr.
This histogram shows that the majority of MBHBs experience delays of $\mathcal{O}(1)$ Gyr; however, a few mergers show delays exceeding the Hubble time.
Consequently, these binaries will not merge within the current cosmic timeline, impacting the anticipated merger rates
and the GW signals detectable by LISA. Further, adding the time delays results in the MBHBs merger occurring at a lower redshift, the mergers that will be detectable with LISA will have an increased SNR. Fig.\ref{fig:SNR_histogram} shows the impact of adding time delays on the SNR of the MBBHs GW signal.

The prolonged time delays can be quantitatively explained through the dynamical friction time scale formula, as referenced in equation \ref{eq:dynamical_friction}. This equation
suggests that extended delays correlate with specific physical parameters: larger initial separations (a) between the black holes, elevated velocity dispersions ($\sigma$)
within the host galaxy, higher ratios of the galaxy mass to the secondary black hole mass ($M_{\rm gal}/M_2$), and reduced masses of the secondary black hole ($M_2$). These factors collectively determine the efficiency and duration of the inspiral process driven by dynamical friction.

\begin{figure}[h]
\centering 
\includegraphics[width=1\linewidth]{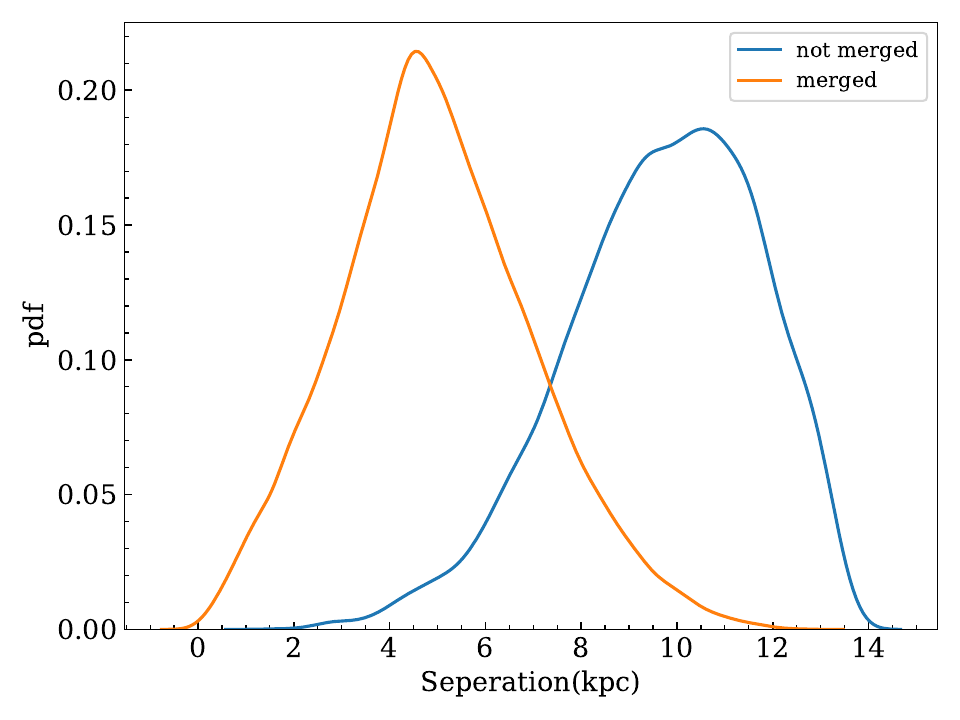}
\caption{Histogram initial separation from the simulation of MBHBs population for merged and not merged MBHBs}
\label{fig:separation_merged_notmerged}
\end{figure}

Further, we find that the likelihood of a MBHB merging
within the Hubble time dominantly hinges on the initial separation of the binary system, rather than other variables such as mass ratios or velocity dispersions.
From Fig \ref{fig:separation_merged_notmerged} we can see that MBHBs with smaller initial separations exhibit a higher merger fraction, as the reduced distance accelerates the dynamical friction process,
enabling coalescence within Hubble time. Conversely, binaries with larger initial separations are more prone to remain unmerged, forming a population
of stalled systems that do not contribute to detectable GW sources. This dependence on initial separation emphasizes the critical
influence of early orbital configurations, as established in simulations, on the evolutionary outcomes of MBHBs.

\subsection{Properties of Host Galaxies of Merging Blackholes}
\begin{figure*}[htp]
\centering
\includegraphics[width = 0.9\textwidth]{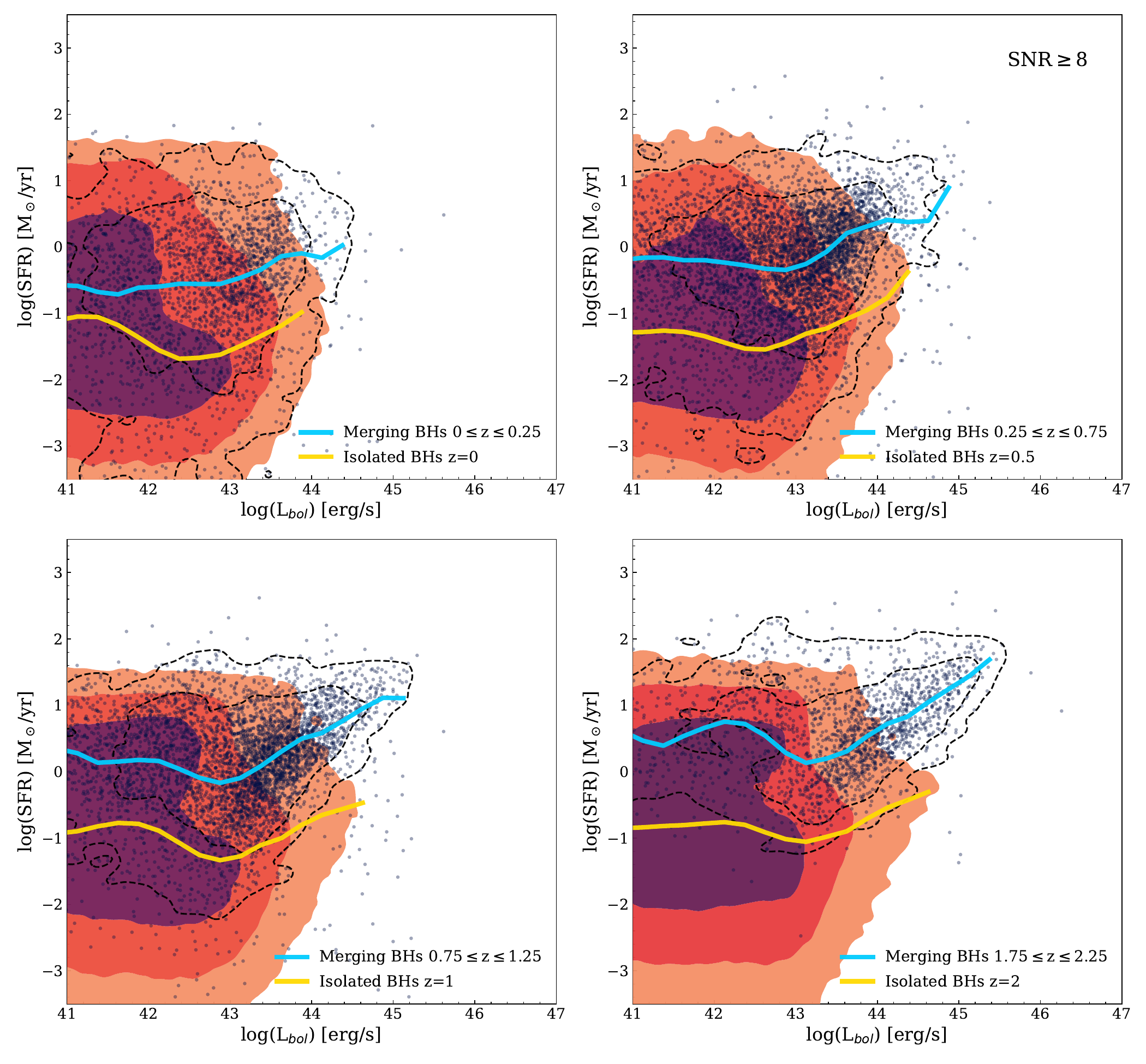}
\caption{A comparison of the SFR of galaxies hosting black holes with $L_{\rm bol}$ at various redshifts.
  The  colored contours represent the 68\%, 95\%, and 99.7\% of the enclosed population for isolated blackholes, and the dashed contours
  represent the 68\%, 95\% of the enclosed population for merging black holes.
  The yellow (cyan) solid line is the median SFR-$L_{\rm bol}$ relation for the isolated (merging) population, and the points represent the
  merging population. The panels represent the scaling relations at redshifts $z=0, 0.5, 1 , 2$.}
\label{fig_sfr-lbol_redshift}
\end{figure*}

In figure~\ref{fig_sfr-lbol_redshift} we look at the SFR-$L_{\rm bol}$ relation of galaxies hosting black holes up to $z=2.$ (different panels).
The colored contours represent the 68\% ($1\sigma$), 95\% (2$\sigma$), and 99.7\% ($3\sigma$) of the enclosed population for isolated blackholes
and the dashed contours represent the $1\sigma$ and $2\sigma$ of the enclosed population for merging black holes. The points represent
data from the merging population only. The solid yellow and cyan lines represent the median relation for the isolated and merging
populations, respectively. The merging population is about 10\% of the isolated population.

The median SFR-$L_{\rm bol}$  relation is flat up to $\log_{10}\left(\Lbol/\Lsun\right) \lesssim 42.5$ and increases thereafter for both the merging and isolated populations.
Comparing the $1\sigma$ and $2\sigma$ populations, we find that the galaxy-black hole relation is tighter for the merging population compared to the isolated population.
Across the full range of luminosities and redshifts explored, we find that the median  SFR at fixed $L_{\rm bol}$ for the merging population is an order of magnitude higher than
the isolated population. However, distinguishing these two populations is hard since they overlap except at the bright end. The largest SFR-$\Lbol$ values predominantly
comes from come from the merging population, even if they represent a small fraction of the overall population. This becomes more obvious at higher redshifts.
The median SFR for the merging population at $\log_{10}(\Lbol/Lsun) \simeq 45 $ increases from $1 \msun/{\rm yr}$ at $z=0$ to $10 \msun/{\rm yr}$ at $z=1$.
At $z=2$ this number is  $100 \msun/{\rm yr}$. We note that the high-SFR bright $\Lbol$ is only probed by the merging population.
Observationally, this regime represents potential GW candidates. However, at this bright end, the quasar outshines the host galaxy, and
a different proxy other than UV would be needed for probing the SFR of the host galaxy. 

\section{Conclusions}
In this work, we make predictions about the detectability of MBHBs in future generation GW detector LISA, which will be detecting GWs of frequency between $10\mu$Hz to $10$mHz. LISA is expected to detect the MBHBs in the mass range of $\sim 10^4-10^8 \msun$. We make use of the low redshift runs of \ninja Simulation Suite with $N_{\rm part} = 2\times1008^3 - 2\times 640^3$ particles and evolve down to $z \sim 0$. For the robust estimation of MBHBs distribution in the \ninja runs, we did the convergence study of the BHMF at $z=0$ and found that the BHMF is converged for $\log_{10}(M_{\rm BH}/\msun) \geq 6.5$, which is about an order of magnitude higher than the seed mass. We limit our analysis to $z\leq 2.5$ as the number of mergers is extremely small at high redshift and in the mass range of $\log_{10}(M_{\rm BH}/\msun) \geq 6.5$ as found in the convergence study. 
A crucial ingredient in these simulations and others is the BH merger tree as well as the properties of the host galaxies. 
In this work we have not looked at the convergence of BH merger rates.
We have assumed that if the BHMF is converged above a certain mass scale, both accretion and merger driven growth should be converged which may not be completely true. Additionally we find that assuming this mass threshold as a convergence
criterion indirectly resolves the host galaxy. We will explore these issues, which directly impact the GW signal 
and events since these appear at lower masses, in a future study.

As the dissipation of orbital energy of MBHBs by GWs is dominant when the separation between the MBHs is $\mathcal{O}(10^{-3})$pc, which is beyond the resolution scale of the simulation, we have to post-process the simulation data to account for the dissipative mechanisms from $\mathcal{O}(\rm kpc)$ to $\mathcal{O}(\rm mpc)$. In this work, we assumed that the principal mechanisms for the orbital evolution of MBHs are due to dynamical friction and stellar hardening. We find that incorporating these two dissipative mechanisms causes a significant shift in the merger time of MBHBs: $\mathcal{O}(1)$Gyr.

To model the GW signal, we use the IMRPhenomA model, which is a phenomenological Inspiral-Merger-Ringdown waveform model for non-spinning BHs. We find that adding spin corrections increases the fractional SNR by merely $\sim 1$ percent, and thus non-spinning waveform model is sufficient for estimating SNRs. To have a robust calculation of the SNR of the GW signal, we also include the SNR from the merger and the ringdown part of the signal along with the SNR from the inspiral. This has the effect of boosting the detectability of the MBH merger as the signal can contribute to the SNR even when the inspiral part of the GW signal is out of the LISA band. We have chosen the detectability criterion of MBHBs to be SNR $\geq 8$. 

Before adding the time delays, we find that the MBHBs possessing a total mass greater than  $\sim 10^{8.4}\msun$ fall out of the detection band, LISA, and the distribution of mass ratio as a function of redshift for the detectable MBHBs does not exhibit any significant deviation from that of the total population of MBHBs from the simulation. We also find that the key factor impeding the detectability of GWs by LISA is the total mass of the MBHs rather than the distance to the binaries 
This is because the merging frequency of the more massive BHs lies below LISA's lower frequency threshold. 

After adding the time delays due to dynamical friction and stellar hardening in the post-processing step, we find that the peak in the redshift distribution of the detectable population is shifted to $z=0.1$, as the peak in the redshift distribution without adding time delays was at $z=0.5$. Incorporating these dissipative mechanisms in the dynamical evolution of MBHBs makes the merger event of the binaries shift to later cosmic epochs, which corresponds to lower redshift values. We also find that few binaries show delays exceeding the Hubble time, predominantly due to the initial separation of the binaries. This shows the critical influence of the orbital dynamics given by the simulation on the evolution of MBHBs. 

As the timescale of dynamical friction and stellar hardening can be long $\sim 1$Gyr, we find this to be consistent with \citep{2019ApJ...879..110K,2020MNRAS.498.2219V}. It is possible that during that period of time, the system encounters additional mergers. We calculated such encounters and found them to be $(\sim0.1 \%)$, which is a small number and will not affect the overall detectable population. There are additional effects in the dynamical evolution of the binaries due to the presence of the circumbinary disk \citep{2015MNRAS.448.3603D, 2019ApJ...871...84M, 2020MNRAS.498.2219V} and loss cone depletion \citep{2004AAS...20514902H, 2007ApJ...660..546S} which contribute to the time delays. However, in our work, we have not considered them. Further, our method of adding the time delays directly has limitations; for instance, after adding the time delays, we assume that the mass of the black hole and other properties of the galaxies do not change, which is not really true, considering that the time delays are $\sim 1$Gyr. Also, the time delays depend significantly on the initial condition of the system e.g., initial separation $(a)$ in Eq.\ref{eq:dynamical_friction}, which depends on the resolution of the simulation. Thus, a suite of high-resolution zoom simulations is needed to track the merging black hole dynamics to much smaller length scales. 
\cite{2023A&A...676A...2D} used zoom-in simulations at  $z \sim 4$ to probe scales of order a few 10 parsecs, 
and obtained the GW signal by post-processing time delays in MBHB mergers. 
Alternately, the GW signal has been estimated by integrating post-Newtonian 
orbits of inspiraling MBHBs in galactic scale simulations \cite{2019ApJ...887...35M}. 
Recently, the GW signal and the signal from their
EM counterparts have been explored by implementing time delay models on-the-fly in galaxy simulations 
\citep{2025A&A...701A.232L}. These different complementary approaches need to be further explored and  
formulated as a subgrid prescription to be implemented 
in cosmological simulations to obtain unbiased statistical estimates of the GW signal.
In our work, we only consider time delays to estimate their effects on the SNR and detectability of the GW signal. Time delays increase the SNR but decrease the number of 
events. However, to truly consider its effects, the dynamical processes of the time delays should be incorporated in the simulations directly and self-consistently with other sub-grid processes.  

We also look at the correlation between the SFR and bolometric luminosity for isolated and merging populations up to $z=2$. The SFR and $L_{\rm bol}$ are calculated using their mean and the median values within 20 Myr of the merger. We find a flat relation between the median SFR-$L_{\rm bol}$ up to $\log_{10}\left(\Lbol/\Lsun\right) \lesssim 42.5$ and increases thereafter for merging as well as isolated populations. Further, we find that the median SFR at fixed $L_{\rm bol}$ for the merging population is an order of magnitude higher than the isolated population, and the correlation for the merging population becomes stronger at high redshifts. This signifies that the regime of high-SFR and bright $L_{\rm bol}$ is a probe of the merging population and a potential GW candidate. 

Our work makes predictions about the population distribution of the detectable MBHBs in LISA using the \ninja Simulation Suite. We highlight the dependence of the simulation parameters on the evolutionary outcomes of time delays of MBHBs during the post-processing step of the analysis. From the observational point of view, we show a strong correlation between SFR-$L_{\rm bol}$ for the merging population over the isolated population of MBHBs. This can have future applications in targeting the GW candidates using LISA and open new avenues of searching MBHBs using multimessenger astronomy. 

\section*{Acknowledgements}

We acknowledge the use of the IUCAA LDG cluster Sarathi, the IUCCA HPC cluster Pegasus, and the NISER HPC facilities for
the computational/numerical work. 
This material is
based upon work supported by NSF’s LIGO Laboratory
which is a major facility fully funded by the National
Science Foundation.
NK would like to thank Simeon Bird for the useful clarifications and discussions regarding 
\mpgadget and for fixing bugs that came up during this simulation campaign.

\textit{Software}: \mpgadget \citep{2018zndo...1451799F}, \software{NumPy} \citep{2011CSE....13b..22V}, \software{SciPy} \citep{2020NatMe..17..261V}, \software{astropy} \citep{2013A&A...558A..33A}, \software{Matplotlib} \citep{2007CSE.....9...90H}, \software{jupyter} \citep{2016ppap.book...87K}.

\section*{Data Availability}
The data supporting the findings of this study, including the simulation datasets generated and analyzed herein, are available from the corresponding author upon reasonable request.

\bibliographystyle{mnras}

\bibliography{new_references}

\end{document}